\documentclass[reprint,superscriptaddress,amsmath,amssymb,aps,prx,longbibliography,floatfix]{revtex4-2}

\usepackage{amsthm}
\usepackage{bm}
\usepackage{bbm}
\usepackage{dsfont}
\usepackage{graphicx}
\usepackage{tikz}
\usetikzlibrary{calc,arrows.meta}
\usepackage{subfig}
\usepackage{booktabs}
\usepackage{xcolor}
\definecolor{citationgreen}{RGB}{45,105,75}
\usepackage[bookmarks=false,breaklinks=false,pdfborder={0 0 1},backref=false,colorlinks=true,citecolor=citationgreen]{hyperref}

\usepackage{orcidlink}
\newtheorem{theorem}{Theorem}

\newcommand{\Tr}{\operatorname{Tr}}

\newcommand{\QCQC}{\mathsf{QC\text{-}QC}}
\hypersetup{
  pdftitle={Symmetry-Reduced Optimization of Quantum-Controlled Causal Order for Multiparameter Metrology},
  pdfauthor={Wenjie Wei and Shengshi Pang},
  pdfsubject={PRX Quantum Research Article},
  pdfkeywords={quantum metrology, multiparameter estimation, indefinite causal order, QC-QC, symmetry reduction}
}

\begin{document}

\title{Symmetry-Reduced Optimization of Quantum-Controlled Causal Order for Multiparameter Metrology}

\author{Wenjie Wei\orcidlink{0009-0001-2435-0852}}
\affiliation{School of Physics, Sun Yat-sen University, Guangzhou, Guangdong 510275, China}

\author{Shengshi Pang\orcidlink{0000-0002-6351-539X}}
\email{pangshsh@mail.sysu.edu.cn}
\affiliation{School of Physics, Sun Yat-sen University, Guangzhou, Guangdong 510275, China}
\affiliation{Hefei National Laboratory, University of Science and Technology of China, Hefei 230088, China}

\date{\today}

\begin{abstract}
Indefinite causal order can enhance quantum metrology by exploiting coherence between different query orders. Quantum circuits with quantum control of causal order (QC-QCs) provide a circuit realization of indefinite causal order with dynamic and coherent control over the order of channel queries, making them promising for multiparameter estimation. Yet optimizing QC-QC strategies is challenging because their causal constraints involve many possible query histories. To bridge this gap, for repeated queries of an identical channel, we derive an exact permutation-symmetry reduction of the QC-QC optimization. The reduction preserves the optimal precision bound while replacing an exponentially growing dual-variable representation with one that grows polynomially in the number of queries. Using conic formulations of multiparameter precision bounds, we compare QC-QC with parallel, fixed-order, causal-superposition, and general indefinite-causal-order strategies. For a noisy qubit channel, an achievable two-query QC-QC precision upper bound lies below the Nagaoka--Hayashi lower bound for fixed-order strategies, demonstrating a strict finite-resource precision advantage by indefinite causal order.
\end{abstract}

\maketitle

\section{Introduction}
\label{sec:introduction}

A main goal of quantum metrology is to establish the
ultimate precision limits for estimating unknown physical
parameters under given resource constraints and to develop
quantum protocols that attain or approach these limits.
By exploiting quantum resources such as coherence and
entanglement, quantum metrology offers the possibility of
achieving precision beyond that attainable by classical
strategies~\cite{giovannetti2004QuantumenhancedMeasurementsBeating,giovannetti2011AdvancesQuantumMetrology,demkowicz-dobrzanski2014UsingEntanglementNoise,pezze2018QuantumMetrologyNonclassical}.
While considerable progress has been made in single-parameter
quantum metrology~\cite{helstrom1976QuantumDetection,braunstein1994StatisticalDistance,fujiwara2008FibreBundle,escher2011GeneralFramework,demkowicz2012ElusiveHeisenberg,kolodynski2013EfficientTools,arrad2014IncreasingSensing,kessler2014QuantumErrorCorrection,demkowicz2017AdaptiveMarkovian,zhou2018AchievingHeisenberg,zhou2020OptimalApproximateQEC,zhou2021AsymptoticTheory,kurdzialek2023UsingAdaptivenessCausal,wei2026FundamentalLimitsQuantum,liu2023OptimalStrategiesQuantum}, many physical systems and practical sensing
tasks inherently involve multiple unknown parameters. Examples include the estimation of different components of a
vector field~\cite{kaubruegger2023OptimalVariationalMultiparameter}, the simultaneous characterization of a physical signal and its associated noise~\cite{vidrighin2014JointEstimationPhase,crowley2014TradeoffSimultaneousQuantum}, and distributed quantum
sensing, where parameter information is encoded across
different sensing
nodes~\cite{gessner2020MultiparameterSqueezingOptimal,proctor2018MultiparameterNetworkedSensors,eldredge2018OptimalSecureSensorNetworks,pezze2025DistributedQuantumMultiparameter,zhang2026DistributedMultiparameterSuperconducting}.
Deriving the ultimate precision of joint parameter
estimation and identifying the quantum resources that can
enhance this precision are thus central problems in
multiparameter quantum metrology~\cite{gorecki2020OptimalProbes,albarelli2022ProbeIncompatibility}.

\begin{figure}[tp]
\centering
\includegraphics[width=0.48\textwidth]{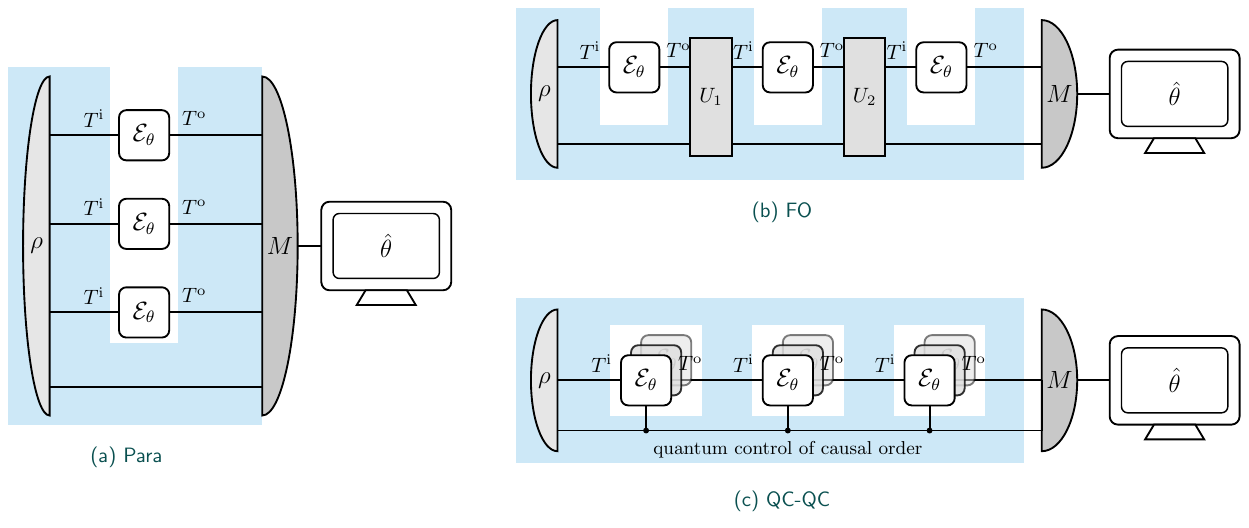}
\caption{Three typical ways to connect channel queries.
(a) Parallel strategy (\(\mathsf{Para}\)). The channels act on a
multipartite probe, and an ancillary system bypasses the queries.
The final measurement can act on all outputs and the ancilla.
(b) Fixed-order strategy (\(\mathsf{FO}\)). Adaptive controls act on
the probe and memory between successive queries in a predetermined
order.
(c) QC-QC strategy. A quantum circuit controls the order of
the queries dynamically and coherently.}
\label{fig:strategy_schematic}
\end{figure}

When the unknown parameters are encoded in a quantum channel,
the attainable precision depends not only on the preparation
of probe states and the choice of measurements but also on
the way in which a given number of channel queries are
connected~\cite{yousefjani2017EstimatingPhaseRandom}.
As illustrated in Fig.~\ref{fig:strategy_schematic},
a parallel strategy prepares all channel inputs before the
queries, allowing the use of entangled multipartite probes,
whereas a fixed-order sequential strategy allows adaptive
quantum controls between successive queries in a predetermined
causal order~\cite{chiribella2009TheoreticalFrameworkQuantum,yuan2016SequentialFeedbackSchemea,kurdzialek2025QuantumMetrologyUsing}.
Beyond these conventional strategies, indefinite causal order (ICO) provides more general ways of
connecting quantum operations
~\cite{oreshkov2012QuantumCorrelationsNo,chiribella2013QuantumComputationsDefinite}.
A prominent example is the quantum switch, which allows
different prescribed orders of quantum operations to be
coherently superposed
~\cite{araujo2015WitnessingCausalNonseparability,wechs2019DefinitionCharacterisationMultipartite}.
More generally, quantum circuits with quantum control of
causal order (QC-QCs) allow the order of subsequent queries
to be determined dynamically and coherently during the
protocol~\cite{wechs2021QuantumCircuitsClassical,grothus2025RoutingQuantumControl,salzger2026HigherOrderQuantumProcesses}.
As the most general class of ICO processes known to admit
circuit realizations without postselection, QC-QCs delimit
the metrological advantages achievable with such realizations
~\cite{mothe2025CorrelationsQuantumCircuits,salzger2026HigherOrderQuantumProcesses}.
ICO has recently emerged as a resource for quantum metrology,
with precision enhancements demonstrated in various scenarios.
The attainable advantage, however, depends on the channel
model, the estimation task, and the precision benchmarks~\cite{zhao2020QuantumMetrologyIndefinite,yin2023ExperimentalSuperHeisenberg,kurdzialek2023UsingAdaptivenessCausal,wei2026FundamentalLimitsQuantum,chapeau-blondeau2021NoisyQuantumMetrology,goldberg2023EvadingNoiseMultiparameter,liu2023OptimalStrategiesQuantum,zhou2024StrictHierarchyOptimal}.
These developments raise a fundamental question: to what extent
can quantum control of causal order improve the precision limit of multiparameter estimation for a given number
of channel queries?

Addressing this question requires taking into account the incompatibility among measurements associated with different parameters~\cite{matsumoto2002NewApproach,ragy2016CompatibilityMultiparameterQuantum,albarelli2020PerspectiveMultiparameterQuantum}.
Unlike in single-parameter estimation, measurements that
are optimal for estimating individual parameters need not
commute with one another. In particular, the symmetric logarithmic derivative (SLD)
Cram\'er--Rao bound need not be attainable in multiparameter
settings~\cite{szczykulska2016MultiParameterQuantumMetrology,albarelli2020PerspectiveMultiparameterQuantum,pezze2025AdvancesMultiparameterQuantum,ragy2016CompatibilityMultiparameterQuantum,liu2020QuantumFisherInformation,conlon2026DecidingAttainabilityMultiparameter}.
When independent copies of the output state are available and
collective measurements across these copies are allowed, the
Holevo--Nagaoka (HN) bound characterizes the asymptotic precision
limit of multiparameter estimation~\cite{holevo2011ProbabilisticStatisticalAspects,yamagata2013QuantumLocalAsymptotic,suzuki2016ExplicitFormulaHolevo,sidhu2021TightBoundsSimultaneousa,albarelli2019EvaluatingHolevoCramerRao}.
In contrast, for a single protocol run involving $N$ channel queries, the
Nagaoka--Hayashi (NH) bound provides a potentially tighter lower
bound on the weighted mean-square error (WMSE) than the HN bound, even when arbitrary joint
measurements on the terminal quantum register are
allowed~\cite{conlon2021EfficientComputationNagaoka,hayashi2023TightCramerRaoType,das2025HolevoWithoutEntangling,conlon2022GapPersistence,tanggara2026OptimalNoisySequential}.
The tight Cram\'er--Rao (TCR) bound, in turn, is defined as the
infimum of the WMSE over all physically realizable measurements on
this register and locally unbiased estimators.
A meaningful comparison of different causal-order strategies
must therefore account for both the strategies
of channel queries and the measurement schemes available
for extracting the encoded parameter information.

Recent conic formulations provide a unified framework for
multiparameter quantum channel estimation, accommodating
different classes of estimation strategies and
different precision bounds
~\cite{zhou2026UnifiedComputableApproach,li2026OptimalStrategiesMultiparameter,hayashi2026MultiparameterQuantumMetrology,tanggara2026OptimalNoisySequential}.
However, optimization over the QC-QC strategy class remains
computationally challenging. Its causal constraints
involve operators associated with subsets of completed queries,
the number of such subsets grows exponentially with the number of
channel slots~\cite{wechs2021QuantumCircuitsClassical}.
This exponential growth limits the applicability of these
formulations to protocols involving many channel queries.

In this work, we exploit permutation symmetry to reduce the
computational complexity of multiparameter quantum channel
estimation with QC-QCs.
Building on the symmetry reduction techniques in semidefinite
optimization and finite-copy quantum estimation
~\cite{bachoc2012InvariantSemidefinitePrograms,yoshida2023ReversingUnknownQubitUnitary,grinko2025MixedSchurWeyl,bergh2026PermutationInvariantOptimization,zhang2026SchurWeylReductionFiniteCopya},
we extend permutation symmetry to the full family of dual history operators and reformulate the recursive causal constraints
in Schur--Weyl blocks.
For repeated queries of the same channel, this reduces the size of the QC-QC dual-variable representation from exponential to polynomial in the number of queries, at fixed channel input and output dimensions and a fixed number of estimated parameters, while preserving the optimal value of the dual problem.
Using this framework, we compare parallel, fixed-order,
causal-superposition, QC-QC, and general ICO strategies
across four noisy qubit-channel models, considering both
the asymptotic HN bound and the finite-resource NH and
adaptive-polytope (AP) formulations of the TCR bound
~\cite{zhou2026UnifiedComputableApproach,li2026OptimalStrategiesMultiparameter,hayashi2026MultiparameterQuantumMetrology,tanggara2026OptimalNoisySequential}.
In particular, for two queries of the amplitude-damping
channel with background dephasing, we obtain an achievable
QC-QC WMSE below the fixed-order NH bound, thereby certifying
a finite-resource precision advantage of quantum control
of causal order in multiparameter channel estimation.

The paper is organized as follows.
Section~\ref{sec:channel-estimation} introduces the
multiparameter channel estimation problem and the causal
strategy classes.
Section~\ref{sec:precision} develops a common conic
formulation for comparing precision bounds across these
strategy classes.
Section~\ref{sec:symmetry} derives the exact symmetry
reduction of the QC-QC dual problem.
Section~\ref{sec:numerics} presents the numerical results,
and Sec.~\ref{sec:conclusion} concludes the paper.

\section{Channel Estimation Problem}
\label{sec:channel-estimation}

In this section, we formulate the multiparameter quantum
channel estimation problem in terms of process measurements
and locally unbiased estimators, introduce the precision
bounds, and characterize the causal strategy classes for
multiple channel queries.

\subsection{Channel model and process measurements}

Let \(\{\mathcal E_\theta\}_{\theta\in\Theta}\) be a differentiable
family of completely positive trace-preserving maps from
\(\mathcal L(\mathcal H_{\rm in})\) to
\(\mathcal L(\mathcal H_{\rm out})\), with
\(\theta=(\theta^1,\ldots,\theta^t)^T\in\Theta\subset\mathbb R^t\).
We take both local dimensions to be \(d\). In a fixed input basis, the
unnormalized Choi operator is
\(\mathsf E_\theta=(\mathcal I_R\otimes\mathcal E_\theta)
(|\mathds{1}\rangle\langle\mathds{1}|)\), with
\(|\mathds{1}\rangle=\sum_a|a\rangle_R\otimes|a\rangle_{\rm in}\)
and \(\mathcal H_R\simeq\mathcal H_{\rm in}\).
It obeys \(\mathsf E_\theta\succeq0\) and
\(\Tr_{\rm out}\mathsf E_\theta=I_{\rm in}\), where we identify \(R\)
with the input slot.

One run of the protocol uses \(N\) instances of the same channel.
We refer to the composite quantum system retained for the final
measurement as the terminal quantum register and its state as the
output state.
Label the input and output of slot \(n\) by \(T^n_{\rm in}\) and
\(T^n_{\rm out}\), and denote their joint space over all slots by
\(\mathcal H_{\rm sys}=\bigotimes_{n=1}^N
(\mathcal H_{T^n_{\rm in}}\otimes\mathcal H_{T^n_{\rm out}})\).
The product \(W_\theta=\mathsf E_\theta^{\otimes N}\) describes the
available black boxes. The order in which the protocol queries them
enters through the tester. As in the process-side formulation of
Ref.~\cite{hayashi2026MultiparameterQuantumMetrology}, the local model
depends on \(W_\theta\) and its tangents. For identical queries,
\begin{equation}
\partial_j W_\theta
=\sum_{n=1}^N
\mathsf E_\theta^{\otimes(n-1)}
\otimes(\partial_j\mathsf E_\theta)
\otimes\mathsf E_\theta^{\otimes(N-n)}.
\label{eq:channel-product-tangents}
\end{equation}
Both \(W_\theta\) and \(\partial_j W_\theta\) are invariant under
relabeling the slots.

A process positive-operator-valued measure (process POVM), also
called a quantum tester, describes the full quantum part of the
protocol on \(\mathcal H_{\rm sys}\). Its effects \(P_x\) incorporate
state preparation, any inter-query controls, and the final measurement.
With the Choi convention above we absorb the tester transpose into
the \(P_x\) and use the trace pairing
\(p_\theta(x)=\Tr[W_\theta P_x]\)
~\cite{chiribella2009TheoreticalFrameworkQuantum,zhou2026UnifiedComputableApproach,li2026OptimalStrategiesMultiparameter,hayashi2026MultiparameterQuantumMetrology}.
The sum \(P=\sum_xP_x\) gives the deterministic tester. In particular,
its pairing with a product of Choi operators equals one for any
trace-preserving channel inserted in each slot, including independent
choices across the slots. For a strategy class \(\mathsf{Strat}\), the
tester satisfies
\begin{equation}
P_x\succeq0,\qquad
P\in\mathcal W_{\mathsf{Strat}},\qquad
\Tr P=d^N.
\label{eq:strategy_constraints}
\end{equation}
Here \(\mathcal W_{\mathsf{Strat}}\) is the unnormalized closed convex
cone of admissible deterministic testers. The trace condition fixes
the scale because the Choi operator of the completely depolarizing
channel is \(I/d\).

\subsection{Local unbiasedness and precision}

The generalized estimator \(\hat P=(\{P_x\}_{x\in\mathcal X},\hat\theta)\)
returns a vector \(\hat\theta(x)\in\mathbb R^t\) for outcome \(x\).
We evaluate it at a specified operating point \(\theta\). Local
unbiasedness requires the correct mean there and unit first-order
response to small changes of the parameters
~\cite{zhou2026UnifiedComputableApproach,li2026OptimalStrategiesMultiparameter,tanggara2026OptimalNoisySequential,zhang2026SchurWeylReductionFiniteCopya}.
Writing \(u_x=\hat\theta(x)-\theta\), these conditions are
\begin{equation}
\begin{aligned}
\sum_x u_x^i\,\Tr[W_\theta P_x]&=0,\\
\sum_x u_x^i\,\Tr[(\partial_jW_\theta)P_x]&=\delta_{ij},
\qquad i,j=1,\ldots,t .
\end{aligned}
\label{eq:channel-local-unbiasedness}
\end{equation}
In the derivative condition, the tester and the numerical values
assigned to its outcomes are held fixed at the operating point. If
only the derivative condition is imposed, subtracting the constant
\(b=\sum_x p_\theta(x)u_x\) from every outcome estimate enforces
the mean condition. Normalization gives
\(\sum_x\partial_jp_\theta(x)=0\), so this shift preserves each
derivative condition.

The mean-square error matrix is
\(V_\theta[\hat P]=\sum_x p_\theta(x)u_xu_x^T\).
It equals the covariance matrix for a locally unbiased estimator.
For a real symmetric weight \(G\succeq0\), the loss is
\begin{equation}
\Tr[GV_\theta[\hat P]]=\sum_x p_\theta(x)\,u_x^TGu_x .
\label{eq:channel-wmse}
\end{equation}
The recentering step, which subtracts \(b\) from every outcome
estimate, reduces this loss by \(b^TGb\), which justifies retaining
only the derivative condition in the optimization in
Sec.~\ref{sec:precision}. For a given \(\mathsf{Strat}\) and \(N\), we
write \(C_\theta(G)\) for the infimum of the WMSE over all admissible
locally unbiased generalized estimators. This is the TCR value for a
single run using \(N\) queries, allowing arbitrary joint measurements
on the terminal quantum register. Collective measurements across
independent copies of the output state, obtained by repeating a fixed
protocol independently, give the separate asymptotic setting
associated with the HN bound.

\subsection{Causal structure of multiple queries}

Figure~\ref{fig:strategy_schematic} illustrates three ways to connect the
slots. A parallel strategy (\(\mathsf{Para}\)) can entangle the input
probes with one another and with an ancilla. A fixed-order strategy
(\(\mathsf{FO}\)) can also process the output and a memory register
before supplying the next input. Since the slots contain identical
channels, we can fix the slot labels in advance for this strategy.
A causal-superposition strategy (\(\mathsf{Sup}\)) can retain coherence
between predetermined permutations. A QC-QC
can control the choice of the next unused slot coherently during
the circuit. General ICO strategies (\(\mathsf{Gen}\)) range over
all positive process matrices obeying the process normalization
constraints~\cite{wechs2021QuantumCircuitsClassical,zhou2026UnifiedComputableApproach,hayashi2026MultiparameterQuantumMetrology}.

These strategy cones obey
\begin{equation}
\mathcal W_{\mathsf{Para}}
\subseteq
\mathcal W_{\mathsf{FO}}
\subseteq
\mathcal W_{\mathsf{Sup}}
\subseteq
\mathcal W_{\QCQC}
\subseteq
\mathcal W_{\mathsf{Gen}}.
\label{eq:strategy_hierarchy}
\end{equation}
The cones describe the deterministic testers after tracing out the
terminal quantum register. A purification of such an operator can
be realized within each class considered here, so the decomposition
into effects \(P_x\) accounts for the measurement on that
register~\cite{mothe2024ReassessingAdvantageIndefinite}.
With one query, all five cones coincide. With two queries and trivial
external past and future systems for the traced process,
\(\mathcal W_{\mathsf{Sup}}=\mathcal W_{\QCQC}\). Larger query counts
allow a strict separation~\cite{wechs2021QuantumCircuitsClassical}.

\subsection{Strategy cones}
\label{sec:strategy-cones}

All strategy sets below are unnormalized closed convex cones. The normalization condition $\Tr P=d^N$ is imposed separately.

\paragraph*{Parallel strategies.}

Let $\mathcal{H}^{(N)}_{{\rm in}}:=\bigotimes^{N}_{i=1}\mathcal{H}_{T^{i}_{{\rm in}}}$
and $\mathcal{H}^{(N)}_{{\rm out}}:=\bigotimes^{N}_{i=1}\mathcal{H}_{T^{i}_{{\rm out}}}$.
The parallel-strategy cone is described by the initial probe state:
\begin{equation}
\mathcal{W}_{\mathsf{Para}}=\left\{ P\in\mathcal{L}(\mathcal{H}_{{\rm sys}})_{+}\mid P=\rho\otimes I_{{\rm out}},\rho\in\mathcal{L}(\mathcal{H}^{(N)}_{{\rm in}})_{+}\right\} .\label{eq:cone_para}
\end{equation}

\paragraph*{Fixed-order strategies.}

First consider the natural order $e=(1,2,\ldots,N)$. The fixed-order
cone is characterized as follows: $P\in\mathcal{W}_{\mathsf{FO}}:=\mathcal{W}^{(e)}_{\mathsf{FO}}$
if and only if there exist positive semidefinite operators $P^{(n)}\in\mathcal{L}(\bigotimes^{n-1}_{i=1}(\mathcal{H}_{T^{i}_{{\rm in}}}\otimes\mathcal{H}_{T^{i}_{{\rm out}}})\otimes\mathcal{H}_{T^{n}_{{\rm in}}})$,
$n=1,\ldots,N$, such that 
\begin{equation}
\begin{aligned}
P&=P^{(N)}\otimes I_{T^N_{\rm out}},\\
\Tr_{T^{n+1}_{\rm in}}P^{(n+1)}
  &=P^{(n)}\otimes I_{T^n_{\rm out}},\\
&\hspace{1em}n=1,\ldots,N-1.
\end{aligned}
\label{eq:cone_fo}
\end{equation}

\paragraph*{Causal-superposition strategies.}

Let $\mathfrak S_N$ denote the permutation group on the $N$
query labels, and let $\mathcal Q_{\mathsf{FO}}^{(\pi)}$
denote the fixed-order constraint map for the order $\pi$,
defined in Appendix~\ref{app:strategy-cones}.
This cone is defined as the Minkowski sum of all fixed-order cones: 
\begin{equation}
\mathcal W_{\mathsf{Sup}}=\left\{
\sum_{\pi\in\mathfrak S_N}Y_\pi\ \middle|\
\begin{gathered}Y_\pi\succeq0,\\
\mathcal Q_{\mathsf{FO}}^{(\pi)}(Y_\pi)=0
\end{gathered}\right\}.
\label{eq:cone_sup}
\end{equation}

\paragraph*{QC-QC strategies.}

We next consider the strategy class that allows dynamic quantum control
of the causal order. Let $\mathcal{N}:=\{1,\ldots,N\}$. We have $P\in\mathcal{W}_{\mathsf{QC\text{-}QC}}$
if and only if there exists a family of positive semidefinite auxiliary
operators $W_{(\mathcal{K}_{n-1},k_{n})}\in\mathcal{L}(\bigotimes_{j\in\mathcal{K}_{n-1}}(\mathcal{H}_{T^{j}_{{\rm in}}}\otimes\mathcal{H}_{T^{j}_{{\rm out}}})\otimes\mathcal{H}_{T^{k_{n}}_{{\rm in}}})_{+}$,
with $n=1,\ldots,N$, $|\mathcal{K}_{n-1}|=n-1$, and $k_{n}\in\mathcal{N}\setminus\mathcal{K}_{n-1}$,
satisfying 
\begin{equation}
P=\sum_{k_{N}\in\mathcal{N}}W_{(\mathcal{N}\setminus\{k_{N}\},k_{N})}\otimes I_{T^{k_{N}}_{{\rm out}}},\label{eq:cone_qcqc_top}
\end{equation}
and, for every $n=1,\ldots,N-1$ and every $\mathcal{K}_{n}\subseteq\mathcal{N}$
with $|\mathcal{K}_{n}|=n$, 
\begin{equation}
\begin{aligned}
&\sum_{k_{n+1}\in\mathcal N\setminus\mathcal K_n}
\Tr_{T^{k_{n+1}}_{\rm in}}W_{(\mathcal K_n,k_{n+1})}\\
&\quad=\sum_{k_n\in\mathcal K_n}
W_{(\mathcal K_n\setminus\{k_n\},k_n)}\otimes I_{T^{k_n}_{\rm out}}.
\end{aligned}
\label{eq:cone_qcqc_rec}
\end{equation}

\paragraph*{General indefinite-causal-order strategies.}

For any subsystem $S$, define the trace-replacement map
\begin{equation}
\mathcal{D}_{S}(X):=\frac{I_{S}}{d_{S}}\otimes\Tr_{S}X.\label{eq:trace_replacement_map}
\end{equation}
When the subscript is omitted, the map denotes complete depolarization on the whole system.

Finally, the cone of general process matrices~\cite{araujo2015WitnessingCausalNonseparability,hoffreumon2023CharacterizationHigherorderQuantum} is $\mathcal{W}_{\mathsf{Gen}}=\left\{ P\succeq0\ \middle|\ \mathcal{Q}_{\mathsf{Gen}}(P)=0\right\} $,
where 
\begin{equation}
\mathcal{Q}_{\mathsf{Gen}}=\bigotimes^{N}_{i=1}\left(\mathcal{I}-\mathcal{D}_{T^{i}_{{\rm out}}}+\mathcal{D}_{T^{i}_{{\rm in}}T^{i}_{{\rm out}}}\right)-\mathcal{D}.\label{eq:qico_appendix}
\end{equation}
The dual cones are given in Appendix~\ref{app:strategy-cones}.

\section{Precision-bound formulation for strategy comparison}
\label{sec:precision}

In this section, we formulate the SLD, HN, NH, and TCR bounds for different causal strategy classes within a unified conic optimization framework. The precision-bound cone specifies the bound under consideration, while the strategy cone specifies the admissible causal structure.

For strategy comparison, we use the established conic construction~\cite{hayashi2023TightCramerRaoType,
hayashi2024FindingOptimalProbe}.
Let
\(\mathcal R\simeq\mathbb C^{t+1}\), with basis
\(\{|0\rangle,|1\rangle,\ldots,|t\rangle\}\). When \(G\) appears as an operator on \(\mathcal R\), we identify it with
\(0\oplus G\), so that \(G|0\rangle=0\). 
Encoding the estimation error for each outcome into a real vector $|\psi_x\rangle = |0\rangle + \sum_{i=1}^t (\hat\theta^i(x)-\theta^i)|i\rangle$, we construct an augmented matrix:
\begin{equation}
X(\hat P) = \sum_x |\psi_x\rangle \langle\psi_x| \otimes P_x.
\label{eq:augmented_variable}
\end{equation}
For \(X\in\mathcal L(\mathcal R\otimes\mathcal H_{\rm sys})\), write
\(X^{j,k}=\langle j|X|k\rangle_\mathcal{R}\).
This compact variable $X$ incorporates the entire estimation protocol: its top-left block $X^{0,0}$ recovers the deterministic process $P$, its off-diagonal blocks $X^{0,i}$ encode the local-unbiasedness constraints $\Tr[\partial_jW_\theta X^{0,i}] = \delta_{ij}$, and its inner product with $(G\otimes W_\theta)$ reproduces \(\Tr[GV_\theta[\hat P]]\). Because the estimator functions are real, $X$ must reside in the real-block linear space 
\begin{equation}
  \label{eq:real_block_space}
  \mathcal B := \{ X=X^\dagger \in \mathcal L( \mathcal R\otimes\mathcal H_{\rm sys} ) \mid (X^{j,k})^\dagger=X^{j,k}, \forall j,k \}.
\end{equation}
Furthermore, the positivity $P_x \succeq 0$ implies that $X$ belongs to $\mathcal S_{\rm SEP}(\mathcal R:\mathcal H_{\rm sys})$, the cone of unnormalized separable states on $\mathcal R\otimes\mathcal H_{\rm sys}$. Thus, the TCR bound is captured by the intersection cone $\mathcal S_{\rm TCR} := \mathcal S_{\rm SEP}(\mathcal R:\mathcal H_{\rm sys}) \cap \mathcal B$.

\begin{figure*}[tp]
\centering
\includegraphics[width=1.7\columnwidth]{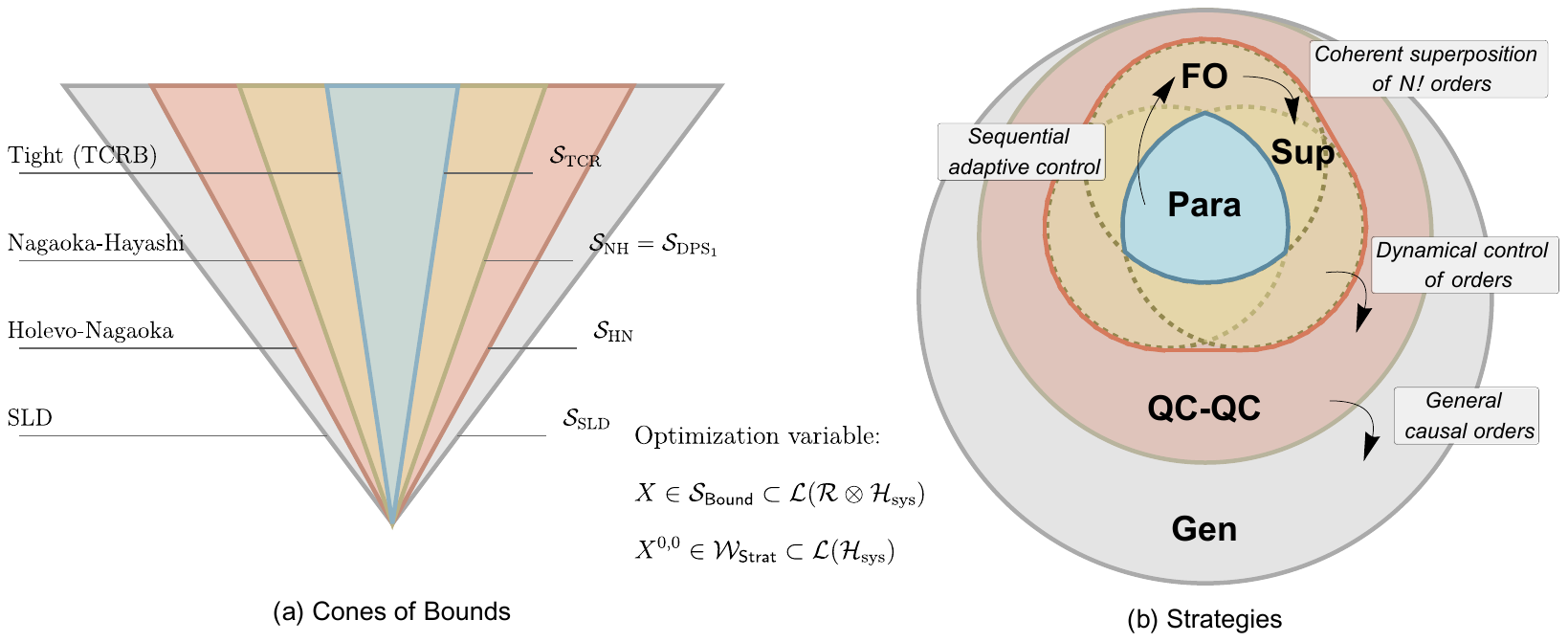}
\caption{Unified conic formulation for comparing strategies and precision
bounds. (a) Hierarchy of precision-bound cones
\(\mathcal{S}_{\mathsf{Bd}}\) on \(\mathcal{R}\otimes\mathcal{H}_{\rm sys}\).
From innermost to outermost, the cones correspond to the tight
Cram\'er--Rao (TCR), Nagaoka--Hayashi (NH), Holevo (HN), and SLD
bounds; the labels Tight (TCRB) and Holevo--Nagaoka in the artwork
refer to the TCR and HN bounds, respectively.
(b) Hierarchy of query strategy cones
\(\mathcal{W}_{\mathsf{Strat}}\) on \(\mathcal{H}_{\rm sys}\). The global
augmented variable \(X\) is constrained by the chosen bound cone
\(\mathcal{S}_{\mathsf{Bd}}\), while its block \(X^{0,0}\) must belong to
the strategy cone \(\mathcal{W}_{\mathsf{Strat}}\).}
\label{fig:cones_and_sets}
\end{figure*}

However, the TCR bound requires optimization over the separable cone, and separability is difficult to decide
~\cite{doherty2004CompleteFamilySeparability}.
If the separable cone is relaxed to the semidefinite cone with suitable linear constraints, one obtains efficiently computable bounds of metrological interest, including the SLD, HN, and NH bounds
~\cite{hayashi2023TightCramerRaoType}.
Together with the unrelaxed TCR cone, we denote these precision-bound cones by
\(\mathcal S_{\mathsf{Bd}}\), where
\(\mathsf{Bd}\in\{\mathsf{SLD},\mathsf{HN},\mathsf{NH},\mathsf{TCR}\}\).
We denote the primal and dual problems by
$P_{\mathsf{Bd}}^{\mathsf{Strat}}$ and
$D_{\mathsf{Bd}}^{\mathsf{Strat}}$, respectively, and write
$S(\cdot)$ for the optimal value of an optimization problem.
With the precision-bound cones and strategy constraints encoded
separately, the relevant optimization problems take the common conic form:
\begin{equation}
\begin{aligned}
S(P_{\mathsf{Bd}}^{\mathsf{Strat}})
=
\inf_X\;
&
\Tr[
(G\otimes W_\theta)X
]
\\
\mathrm{s.t.}\;
&
\Tr[
\partial_jW_\theta X^{0,i}
]
=
\delta_{ij},
\;
X\in\mathcal S_{\mathsf{Bd}},
\\
&
X^{0,0}
\in
\mathcal W_{\mathsf{Strat}},
\;
\Tr X^{0,0}
=
d^{N}.
\end{aligned}
\label{eq:unified_program}
\end{equation}
Equation~\eqref{eq:unified_program} separates the strategy cone
$\mathcal W_{\mathsf{Strat}}$ from the precision-bound cone
$\mathcal S_{\mathsf{Bd}}$, as illustrated in Fig.~\ref{fig:cones_and_sets}.
Let $R_{ab}={\rm i}(|a\rangle\langle b|-|b\rangle\langle a|)$.
Define the linear space associated with the SLD bound, 
\begin{equation}
\mathcal B''=\left\{X=X^\dagger\ \middle|\
\begin{gathered}
X\in\mathcal L(\mathcal R\otimes\mathcal H_{\rm sys}),\\
(X^{0,k})^\dagger=X^{0,k},\ k=1,\ldots,t
\end{gathered}\right\}.
\label{eq:sm_B_double_prime}
\end{equation}
The bound cones for the four precision bounds in the unified primal
problem are 
\begin{equation}
\begin{aligned}
\mathcal S_{\mathsf{SLD}}&=\mathcal L(\mathcal R\otimes\mathcal H_{\rm sys})_+\cap\mathcal B'',\\
\mathcal S_{\mathsf{HN}}&=\left\{X\in\mathcal S_{\mathsf{SLD}}\ \middle|\
\begin{gathered}
\Tr[X(R_{jk}\otimes W_\theta)]=0,\\
1\le j<k\le t
\end{gathered}\right\},\\
\mathcal S_{\mathsf{NH}}&=\mathcal L(\mathcal R\otimes\mathcal H_{\rm sys})_+\cap\mathcal B,\\
\mathcal S_{\mathsf{TCR}}&=\mathcal S_{\rm SEP}(\mathcal R:\mathcal H_{\rm sys})\cap\mathcal B.
\end{aligned}
\label{eq:sm_bound_cones}
\end{equation}
Here $\mathcal{S}_{{\rm SEP}}(\mathcal{R}:\mathcal{H}_{{\rm sys}})$
is the cone of unnormalized separable operators between the auxiliary
parameter space and the physical process space. The cone definitions
in Eq.~\eqref{eq:sm_bound_cones} give the inclusion chain~\cite{hayashi2023TightCramerRaoType}
$\mathcal{S}_{\mathsf{TCR}}\subseteq\mathcal{S}_{\mathsf{NH}}\subseteq\mathcal{S}_{\mathsf{HN}}\subseteq\mathcal{S}_{\mathsf{SLD}}.$

\subsection{TCR conic program for general strategies}
\label{sec:tcr-recovery}
For any strategy class $\mathsf{Strat}$, setting $\mathsf{Bd}=\mathsf{TCR}$ in Eq.~\eqref{eq:unified_program} gives an exact conic formulation of the locally unbiased estimation problem.
Every generalized estimator produces a feasible $X(\hat P)$ through Eq.~\eqref{eq:augmented_variable}, with objective value equal to its WMSE. Thus
\begin{equation}
S(P_{\mathsf{TCR}}^{\mathsf{Strat}})
\le C_\theta(G).
\label{eq:tcr_lower_direction}
\end{equation}
Differentiating the process normalization gives
\begin{equation}
\Tr[\partial_jW_\theta P]=0.
\label{eq:process_normalization_derivative}
\end{equation}
We next prove the reverse inequality. Take any feasible point $X\in\mathcal{S}_{{\rm SEP}}(\mathcal{R}:\mathcal{H}_{{\rm sys}})\cap\mathcal{B}$
of $(P^{\mathsf{Strat}}_{\mathsf{TCR}})$. Because $X$ lies in the
separable cone, and because $X\in\mathcal{B}$ implies $X^{T_{\mathcal{R}}}=X$,
we can average an arbitrary separable decomposition over partial transposition
on the auxiliary space and then take a real spectral decomposition
of the auxiliary factors. Hence ${\displaystyle X=\sum^{m}_{\alpha=1}T_{\alpha}\otimes M_{\alpha}}$,
where each $T_{\alpha}$ is a rank-one real symmetric positive semidefinite
matrix on the auxiliary space and $M_{\alpha}\succeq0$. Let $\mathcal{K}_{v}=\{\alpha:\langle0|T_{\alpha}|0\rangle>0\}$,
and define ${\displaystyle X_{v}=\sum_{\alpha\in\mathcal{K}_{v}}T_{\alpha}\otimes M_{\alpha}.}$
When $\langle0|T_{\alpha}|0\rangle=0$ we have $T_{\alpha}|0\rangle=0$,
since $T_{\alpha}\succeq0$, and therefore $\langle0|T_{\alpha}|i\rangle=0$.
Thus $X^{0,0}_{v}=X^{0,0},X^{0,i}_{v}=X^{0,i}.$ These block identities
imply $X^{0,0}_{v}\in\mathcal{W}_{\mathsf{Strat}}$, $\operatorname{Tr}X^{0,0}_{v}=d^N$,
and 
\begin{equation}
\operatorname{Tr}\left[\partial_{j}W_{\theta}X^{0,i}_{v}\right]=\delta_{ij}.\label{eq:TCR_valid_unbiased_constraint}
\end{equation}
Moreover, $X-X_{v}\succeq0$ and $G\otimes W_{\theta}\succeq0$, so
\begin{equation}
\operatorname{Tr}\left[(G\otimes W_{\theta})X_{v}\right]\leq\operatorname{Tr}\left[(G\otimes W_{\theta})X\right].\label{eq:TCR_valid_objective_monotonicity}
\end{equation}
We now construct a generalized estimator $\hat{P}_{X}:=(\{P_{\alpha}\},\hat{\theta})$
from $X_{v}$. For each $\alpha\in\mathcal{K}_{v}$, define 
\begin{equation}
P_{\alpha}=\langle0|T_{\alpha}|0\rangle M_{\alpha},\;\hat{\theta}^{i}(\alpha)=\theta^{i}+\frac{\langle0|T_{\alpha}|i\rangle}{\langle0|T_{\alpha}|0\rangle}.\label{eq:TCR_reconstructed_estimator_general_theta}
\end{equation}
Since $T_{\alpha}$ is real symmetric, $\hat{\theta}^{i}(\alpha)$
is real. Also, 
\begin{equation}
\sum_{\alpha\in\mathcal{K}_{v}}P_{\alpha}=X^{0,0}_{v}=X^{0,0}\in\mathcal{W}_{\mathsf{Strat}}.\label{eq:TCR_reconstructed_process_general_theta}
\end{equation}
Therefore, $\{P_{\alpha}\}_{\alpha\in\mathcal{K}_{v}}$ is a valid
probabilistic process under the given strategy class. The estimator
in Eq.~\eqref{eq:TCR_reconstructed_estimator_general_theta} satisfies
the derivative condition for local unbiasedness. Indeed, Eqs. ~\eqref{eq:process_normalization_derivative},
~\eqref{eq:TCR_valid_unbiased_constraint} and ~\eqref{eq:TCR_reconstructed_process_general_theta}
give 
\begin{equation}
\begin{aligned}
&\sum_{\alpha\in\mathcal K_v}\hat\theta^i(\alpha)
  \Tr[\partial_jW_\theta P_\alpha]\\
&=\sum_{\alpha\in\mathcal K_v}\langle0|T_\alpha|i\rangle
  \Tr[\partial_jW_\theta M_\alpha]\\
&\quad+\theta^i\Tr[\partial_jW_\theta X_v^{0,0}]\\
&=\Tr[\partial_jW_\theta X_v^{0,i}]=\delta_{ij}.
\end{aligned}
\end{equation}

It remains to compare the WMSE of $\hat{P}_{X}$ with the conic-program
objective. Since $T_{\alpha}$ has rank one, the matrix elements of
$T_{\alpha}$ satisfy $\langle0|T_{\alpha}|i\rangle\langle0|T_{\alpha}|j\rangle=\langle0|T_{\alpha}|0\rangle\langle i|T_{\alpha}|j\rangle$.
One then verifies that $\operatorname{Tr}\left[GV_{\theta}[\hat{P}_{X}]\right]=\operatorname{Tr}\left[(G\otimes W_{\theta})X_{v}\right]$.
Combining this identity with Eq. ~\eqref{eq:TCR_valid_objective_monotonicity}
gives 
\begin{equation}
C_{\theta}(G)\leq\operatorname{Tr}\left[GV_{\theta}[\hat{P}_{X}]\right]\leq\operatorname{Tr}\left[(G\otimes W_{\theta})X\right].\label{eq:TCR_reverse_bound_general_theta}
\end{equation}
A constant shift fixes $\sum_{\alpha}\hat\theta^i(\alpha)p_\theta(\alpha)=\theta^i$ without changing the derivative condition and can only decrease the WMSE.

Since Eq. ~\eqref{eq:TCR_reverse_bound_general_theta} holds for
every feasible point $X$ of the conic program, taking the infimum
over all feasible points gives 
\begin{equation}
C_{\theta}(G)\leq S(P^{\mathsf{Strat}}_{\mathsf{TCR}}).\label{eq:TCR_reverse_inequality_general_theta}
\end{equation}

It follows that
\begin{equation}
C_\theta(G)=S(P_{\mathsf{TCR}}^{\mathsf{Strat}}).
\label{eq:tcr_exact}
\end{equation}
The TCR program therefore gives the physical optimum, whereas the SLD, HN, and NH programs are semidefinite-program (SDP) lower-bound relaxations without a general guarantee of finite-query attainability~\cite{conlon2021EfficientComputationNagaoka}.

\subsection{Single-level and two-level equivalence}
\label{sec:single-double}
This unified single-level program also has a direct two-level operational interpretation.
Let \(F\) denote the terminal quantum register retained after all channel queries, with Hilbert space \(\mathcal H_F\), and let
\(\widetilde P\succeq0\)
be a deterministic channel-query strategy on
\(\mathcal H_{\rm sys}\otimes\mathcal H_F\)
that satisfies $\Tr_F\widetilde P
\in
\mathcal W_{\mathsf{Strat}},
$ and $
\Tr\widetilde P
=
d^N$.
Contracting this extended process $\widetilde{P}$ with the channel Choi operator
\(W_\theta\)
gives the output state on \(F\):
\begin{equation}
\rho_\theta(\widetilde P)
=
\Tr_{\rm sys}
\left[
(W_\theta\otimes I_F)
\widetilde P
\right].
\label{eq:effective_output_state}
\end{equation}
For a fixed \(\widetilde P\), the inner level optimizes the precision
bound for the induced output-state model
\(\{\rho_\theta(\widetilde P)\}_{\theta\in\Theta}\).
Let
\(S(P_{\mathsf{Bd}}[
\rho_\theta(\widetilde P)
])\)
denote the optimal precision-bound value for the induced output-state model.
Further optimizing the outer strategy gives
\begin{equation}
\begin{aligned}
K(P_{\mathsf{Bd}}^{\mathsf{Strat}}):=
\inf_{\substack{
\widetilde P\succeq0\\
\Tr_F\widetilde P
\in
\mathcal W_{\mathsf{Strat}}\\
\Tr\widetilde P=
d^N
}}
S(
P_{\mathsf{Bd}}
[
\rho_\theta(\widetilde P)
]
).
\end{aligned}
\label{eq:two_level_program}
\end{equation}
Equation~\eqref{eq:two_level_program} is the original two-level task, while
Eq.~\eqref{eq:unified_program} folds the same strategy and estimation
variables into the single augmented variable \(X\).
\begin{theorem}[Equivalence between the two-level optimization and the single-level conic program]
\label{thm:two_level_single_level_equivalence}
Fix a strategy class $\mathsf{Strat}$, and let
$\mathsf{Bd}\in\{\mathsf{SLD},\mathsf{HN},\mathsf{NH}\}$. Then the two formulations in Eqs.~\eqref{eq:two_level_program} and~\eqref{eq:unified_program} have the same optimal value:
\begin{equation}
K(P^{\mathsf{Strat}}_{\mathsf{Bd}})=S(P^{\mathsf{Strat}}_{\mathsf{Bd}}).
\label{eq:two_level_single_level_equivalence}
\end{equation}
\end{theorem}
This equivalence applies to the three SDP relaxations. The TCR program is treated separately by the recovery result in Eq.~\eqref{eq:tcr_exact}. Thus the single-level program is an exact convex reformulation, which we use for the dual, symmetry reduction, and numerical algorithms below.

\subsection{Dual program and strong duality}
\label{sec:dual}
The dual of Eq.~\eqref{eq:unified_program} exposes the strategy-cone constraints needed for symmetry reduction and numerical optimization.
Let $\Xi_{\mathsf{Bd}}$ be the set of dual variables associated with the precision bound, consisting of $h$ for the local-unbiasedness constraints, $\mu$ for the strategy-normalization constraint, $S$ for the strategy cone, and the remaining components describing the precision-bound relaxation.
Then the dual of~\eqref{eq:unified_program} reads
\begin{equation}
\begin{aligned}
S(D_{\mathsf{Bd}}^{\mathsf{Strat}})
:=
\max_{\Xi_{\mathsf{Bd}}}\;
&
-\Tr h
-
\mu d^{N}
\\
\mathrm{s.t.}\;
&
B_{\mathsf{Bd}}
(
\Xi_{\mathsf{Bd}}
)
\in
\mathcal K_{\mathsf{Bd}}^{*},
\;
\mu I_{\rm sys}
-
S
\in
\mathcal W_{\mathsf{Strat}}^{*}.
\end{aligned}
\label{eq:unified_dual}
\end{equation}
Here
\(B_{\mathsf{Bd}}
(
\Xi_{\mathsf{Bd}}
)\)
collects the cost matrix
\(G\), the channel Choi operator
\(W_\theta\), the parameter derivatives
\(\partial_jW_\theta\)
and the precision-bound dual variables.
The cone
\(\mathcal K_{\mathsf{Bd}}^{*}\)
is the dual cone of the corresponding bound.
For
\(\mathsf{SLD}\), \(\mathsf{HN}\), and \(\mathsf{NH}\),
\(\mathcal K_{\mathsf{Bd}}^{*}\)
is a semidefinite cone, while for
\(\mathsf{TCR}\),
\(\mathcal K_{\mathsf{Bd}}^{*}\)
is the dual cone of the separable cone $\mathcal{S}_{\rm SEP}$.

For example, under the parallel strategy with the NH bound,
\(\Xi_{\mathsf{NH}}\)
can be written as
\((h,\mu,S,\{Z_k\}_{k=1}^{t},\{V_{jk}\}_{1\le j<k\le t})\),
where \(Z_k=Z_k^\dagger\) and \(V_{jk}=V_{jk}^\dagger\) act on
\(\mathcal H_{\rm sys}\)
and are the multipliers for the Hermitian constraints in Eq.~\eqref{eq:real_block_space}.
The multiplier \(h\) enters the precision-bound dual operator through \(L(h)=\frac{1}{2}\sum_{i,j=1}^{t}h_{ij}\left(|0\rangle\langle i|+|i\rangle\langle0|\right)\otimes\partial_jW_\theta\).
Let
\(R_{ab}={\rm i}(|a\rangle\langle b|-|b\rangle\langle a|)\). The precision-bound operator is
\begin{equation}
B_{\mathsf{NH}}
\!=\!
L(h)
+
\sum_{k}
R_{0k}\!\otimes\! Z_k
+
\sum_{j<k}
R_{jk}\!\otimes\! V_{jk}
+
|0\rangle\langle0|\!\otimes\! S
+
G\!\otimes\! W_\theta.
\label{eq:para_nh_measurement_dual}
\end{equation}
The first dual constraint is then
\(B_{\mathsf{NH}}\succeq0\),
while the second constraint for the parallel strategy is equivalent to
\(\mu d^{N}I_{\rm in}-\Tr_{\rm out}S\succeq0\).
For the TCR bound, the first constraint is instead the dual-cone constraint of the separable cone between the auxiliary and system spaces.

The corresponding dual test operators for the other precision bounds are
\begin{equation}
\begin{aligned}B_{\mathsf{SLD}}={}&L(h)+\sum_{k=1}^tR_{0k}\otimes Z_k\\&+|0\rangle\langle0|\otimes S+G\otimes W_\theta.\end{aligned}
\label{eq:B_SLD}
\end{equation}
\begin{equation}
\begin{aligned}
B_{\mathsf{HN}}={}&L(h)+\sum_{k=1}^tR_{0k}\otimes Z_k\\
&+\sum_{1\le j<k\le t}\nu_{jk}R_{jk}\otimes W_\theta\\
&+|0\rangle\langle0|\otimes S+G\otimes W_\theta.
\end{aligned}
\label{eq:B_HN}
\end{equation}
\begin{equation}
\begin{aligned}B_{\mathsf{TCR}}={}&L(h)+\sum_{0\le j<k\le t}R_{jk}\otimes Q_{jk}\\&+|0\rangle\langle0|\otimes S+G\otimes W_\theta.\end{aligned}
\label{eq:B_TCR}
\end{equation}
Here $\nu_{jk}\in\mathbb R$ and $Q_{jk}=Q_{jk}^{\dagger}$. The basic dual cones are
\begin{equation}
\mathcal{K}^{*}_{\mathsf{Bd}}=\begin{cases}
\mathcal{L}(\mathcal{R}\otimes\mathcal{H}_{{\rm sys}})_{+}, & \mathsf{Bd}\in\{\mathsf{SLD},\mathsf{HN},\mathsf{NH}\},\\
\mathcal{S}^{*}_{{\rm SEP}}(\mathcal{R}:\mathcal{H}_{{\rm sys}}), & \mathsf{Bd}=\mathsf{TCR}.
\end{cases}\label{eq:measurement_base_dual_cones}
\end{equation}

If \(\{
\partial_iW_\theta
\}_{i=1}^{t}\)
are linearly independent, the primal problem
~\eqref{eq:unified_program}
satisfies Slater's condition~\cite{boyd2004ConvexOptimization}.
Hence \((P_{\mathsf{Bd}}^{\mathsf{Strat}})\) and \((D_{\mathsf{Bd}}^{\mathsf{Strat}})\) have no duality gap, \(S(P_{\mathsf{Bd}}^{\mathsf{Strat}})=S(D_{\mathsf{Bd}}^{\mathsf{Strat}})\).
The explicit Slater construction and proof of strong duality are given in
Appendix~\ref{app:slater_condition}.

\section{Exact symmetry reduction of QC-QC causal histories}
\label{sec:symmetry}

In this section, we exploit the permutation symmetry of identical channel queries to reduce the QC-QC dual optimization problems without changing the resulting SLD, HN, NH, and TCR precision bounds. After permutation averaging, the
$2^N-1$ dual history operators indexed by nonempty subsets
of query labels are determined by $N$ representative
operators, one for each subset $\{1,\ldots,m\}$ with
$m=1,\ldots,N$. Schur--Weyl decomposition of these
representatives yields a dual-variable representation whose
size grows polynomially with $N$ for fixed $d$ and $t$.

\subsection{History-orbit reduction}

Let $\mathcal N=\{1,\ldots,N\}$ denote the query labels.
For the witnesses $H_{\mathcal K}$ of the strategy-dual constraint, with $H_{\mathcal N}=\mu I_{\rm sys}-S$, set $S_{\mathcal K}=d^{N-|\mathcal K|}\mu I_{\mathcal K}-H_{\mathcal K}$, so that $S_{\mathcal N}=S$.
For every nonempty $\mathcal K\subseteq\mathcal N$, the QC-QC dual contains a history operator $S_{\mathcal K}$, and the causal constraints couple the operator for a completed history to that of a history extended by an additional query:
\begin{equation}
S_{\mathcal K}\otimes I_{T^j_{\rm in}}
-\Tr_{T^j_{\rm out}}S_{\mathcal K\cup\{j\}}\succeq0,
\quad j\notin\mathcal K.
\label{eq:qcqc_history_main}
\end{equation}
The bottom-level constraint is
$d^N\mu I_{T^j_{\rm in}}-\Tr_{T^j_{\rm out}}S_{\{j\}}\succeq0$.

For identical channel queries, the product Choi operator $W_\theta=\mathsf E_\theta^{\otimes N}$ is invariant under permutation of the query labels. For each $\pi\in\mathfrak S_N$, let $U_{\pi,\mathcal K}$ relabel the tensor factors in the history $\mathcal K$. The simultaneous action
\begin{equation}
(\pi\cdot\mathsf S)_{\mathcal K}
=U_{\pi,\mathcal K}
S_{\pi^{-1}(\mathcal K)}
U_{\pi,\mathcal K}^{\dagger}
\label{eq:qcqc_action_main}
\end{equation}
preserves both the causal recursion and the dual objective. Hence a group average of an optimal family of dual history operators remains optimal and feasible, whereas averaging only the top-level operator would break causal consistency across histories.

Let $\mathcal{K}_{m}:=\{1,\ldots,m\}$ and write $S^{(m)}:=S_{\mathcal{K}_{m}}$.
A fixed-point node family is uniquely determined by canonical representatives
satisfying $[S^{(m)},U_{\tau}]=0$ for $\tau\in\mathfrak{S}_{m}$,
with 
\begin{equation}
S_{\mathcal{K}}=U_{\sigma,\mathcal{K}}S^{(m)}U^{\dagger}_{\sigma,\mathcal{K}},\;|\mathcal{K}|=m,\;\sigma(\mathcal{K}_{m})=\mathcal{K}.\label{eq:QCQC_reconstruct_nodes_from_representative_symred}
\end{equation}
Stabilizer invariance makes the right-hand side independent of the
choice of $\sigma$. By permutation covariance, all node recursions
are equivalent to the canonical chain 
\begin{align}
S^{(m-1)}\otimes I_{T^{m}_{{\rm in}}}-\operatorname{Tr}_{T^{m}_{{\rm out}}}S^{(m)} & \succeq0, &  & m=2,\ldots,N,\label{eq:qcqc_canonical_main}\\
d^N\mu I_{T^{1}_{{\rm in}}}-\operatorname{Tr}_{T^{1}_{{\rm out}}}S^{(1)} & \succeq0.\label{eq:QCQC_canonical_bottom_constraint_symred}
\end{align}

\subsection{Schur--Weyl block recursion}

Let $\mathbb Y_m^{(q)}$ denote the set of Young diagrams
with $m$ boxes and at most $q$ rows.
For each $m$, let
$\mathcal{H}^{(m)}:=\bigotimes^{m}_{\ell=1}\left(
\mathcal{H}_{T^{\ell}_{{\rm in}}}\otimes
\mathcal{H}_{T^{\ell}_{{\rm out}}}\right)$.
Let $\mathsf U_m$ denote the Schur transform, which gives
\[
\mathsf U_m\mathcal H^{(m)}=
\bigoplus_{\lambda\in\mathbb Y_m^{(d^2)}}
\mathcal U_\lambda^{(m)}\otimes\mathcal S_\lambda^{(m)}.
\]
Here $\mathcal U_\lambda^{(m)}$ and $\mathcal S_\lambda^{(m)}$
are the corresponding irreducible representation spaces of
$\mathrm U(d^2)$ and $\mathfrak S_m$, respectively.

Since $S^{(m)}$ is $\mathfrak{S}_{m}$-invariant, there exist Hermitian
blocks $s^{(m)}_{\lambda}\in\mathcal{L}(\mathcal{U}^{(m)}_{\lambda})$
such that 
\begin{equation}
\mathsf{U}_{m}S^{(m)}\mathsf{U}^{\dagger}_{m}=\bigoplus_{\lambda\in\mathbb{Y}^{(d^{2})}_{m}}s^{(m)}_{\lambda}\otimes I_{\mathcal{S}^{(m)}_{\lambda}}.\label{eq:schur_weyl}
\end{equation}

We write $\nu\nearrow\lambda$ if $\lambda$ is obtained
from $\nu$ by adding one box.
If $\nu\in\mathbb{Y}^{(d^{2})}_{m-1}$, $\lambda\in\mathbb{Y}^{(d^{2})}_{m}$,
and $\nu\nearrow\lambda$, choose the corresponding Clebsch--Gordan
isometry $\mathcal{V}^{(m)}_{\nu\to\lambda}:\mathcal{U}^{(m)}_{\lambda}\longrightarrow\mathcal{U}^{(m-1)}_{\nu}\otimes\mathcal{H}_{T^{m}_{{\rm in}}}\otimes\mathcal{H}_{T^{m}_{{\rm out}}},$
and define 
\begin{equation}
\Phi^{(m)}_{\lambda\to\nu}(x):=\operatorname{Tr}_{T^{m}_{{\rm out}}}\!\left[\mathcal{V}^{(m)}_{\nu\to\lambda}x\bigl(\mathcal{V}^{(m)}_{\nu\to\lambda}\bigr)^{\dagger}\right].\label{eq:QCQC_reduced_step_map_symred}
\end{equation}
For $m=2,\ldots,N$ and $\nu\in\mathbb Y_{m-1}^{(d^2)}$, the canonical-chain constraints become 
\begin{align}
s^{(m-1)}_{\nu}\otimes I_{T^{m}_{{\rm in}}}-\sum_{\lambda:\,\nu\nearrow\lambda}\Phi^{(m)}_{\lambda\to\nu}\!\left(s^{(m)}_{\lambda}\right) & \succeq0,\label{eq:qcqc_young_recursion_main}\\
d^N\mu I_{T^{1}_{{\rm in}}}-\sum_{\lambda\in\mathbb{Y}^{(d^{2})}_{1}}\Phi^{(1)}_{\lambda\to\varnothing}\!\left(s^{(1)}_{\lambda}\right) & \succeq0,\label{eq:QCQC_reduced_bottom_constraint_symred}
\end{align}
where $\varnothing$ denotes the trivial representation at level zero.

For $m=N$, we abbreviate $\mathsf U_N$,
$\mathcal U_\lambda^{(N)}$, and $\mathcal S_\lambda^{(N)}$
as $\mathsf U$, $\mathcal U_\lambda$, and
$\mathcal S_\lambda$, respectively.

For a permutation-invariant operator $X$, denote by $[X]_{\lambda}$
the $\lambda$-th Schur block determined by 
\begin{equation}
\mathsf{U}X\mathsf{U}^{\dagger}=\bigoplus_{\lambda\in\mathbb{Y}^{(d^{2})}_{N}}[X]_{\lambda}\otimes I_{\mathcal{S}_{\lambda}}.\label{eq:schur_block_extraction_symred}
\end{equation}
Conversely, for a block family $x=\{x_{\lambda}\}_{\lambda\in\mathbb{Y}^{(d^{2})}_{N}}$,
define the corresponding full-space operator 
\begin{equation}
\iota_{{\rm SW}}(x):=\mathsf{U}^{\dagger}\left(\bigoplus_{\lambda\in\mathbb{Y}^{(d^{2})}_{N}}x_{\lambda}\otimes I_{\mathcal{S}_{\lambda}}\right)\mathsf{U}.\label{eq:app_sw_embedding_ico_sup}
\end{equation}
Write $W_{\theta,\lambda}:=[W_{\theta}]_{\lambda}$, $\partial_{j}W_{\theta,\lambda}:=[\partial_{j}W_{\theta}]_{\lambda}$,
and denote the corresponding Schur blocks of Eqs.~\eqref{eq:para_nh_measurement_dual}, \eqref{eq:B_SLD}, \eqref{eq:B_HN}, and~\eqref{eq:B_TCR}
by $B^{\mathsf{SLD}}_{\lambda}$, $B^{\mathsf{HN}}_{\lambda}$, $B^{\mathsf{NH}}_{\lambda}$,
and $B^{\mathsf{TCR}}_{\lambda}$. Thus 
\begin{equation}
B^{\mathsf{Bd}}_{\lambda}\succeq0,\;\lambda\in\mathbb{Y}^{(d^{2})}_{N},\;\mathsf{Bd}\in\{\mathsf{SLD},\mathsf{HN},\mathsf{NH}\}.\label{eq:measurement_block_psd_symred}
\end{equation}
For the TCR bound, define the symmetry-reduced dual separable
cone (where SR stands for symmetry reduced)
\begin{equation}
\begin{aligned}
\mathcal S_{\rm SEP}^{*,{\rm SR}}
  :=\bigl\{(Y_\lambda)_\lambda\ \big|\ &
  (\operatorname{id}_{\mathcal R}\otimes\iota_{\rm SW})
  ((Y_\lambda)_\lambda)\\
  &\in\mathcal S_{\rm SEP}^*\bigr\}.
\end{aligned}
\label{eq:sep_dual_reduced_symred}
\end{equation}
Here the embedding acts on each system block and as the identity on $\mathcal R$.
The precision-bound constraint is 
\begin{equation}
\left(B^{\mathsf{TCR}}_{\lambda}\right)_{\lambda\in\mathbb{Y}^{(d^{2})}_{N}}\in\mathcal{S}^{*,{\rm SR}}_{{\rm SEP}}.\label{eq:tcr_measurement_reduced_symred}
\end{equation}

\begin{figure}[htbp]
\centering \begin{tikzpicture}[line width=0.45pt, scale=0.75]
  \tikzset{
    bratteliarrow/.style={->, >=stealth, shorten >=15pt, shorten <=15pt}
  }

  \coordinate (v00) at (0.0,  0.0);
  \coordinate (v10) at (1.60,  0.0);
  \coordinate (v20) at (3.20,  1.00);
  \coordinate (v21) at (3.20, -1.00);
  \coordinate (v30) at (5.20,  1.80);
  \coordinate (v31) at (5.20,  0.0);
  \coordinate (v32) at (5.20, -1.80);
  \coordinate (v40) at (7.50,  2.40);
  \coordinate (v41) at (7.50,  1.20);
  \coordinate (v42) at (7.50,  0.0);
  \coordinate (v43) at (7.50, -1.20);
  \coordinate (v44) at (7.50, -2.40);

  \node[above] at (0.0, 2.85) {$m=0$};
  \node[above] at (1.60, 2.85) {$m=1$};
  \node[above] at (3.20, 2.85) {$m=2$};
  \node[above] at (5.20, 2.85) {$m=3$};
  \node[above] at (7.50, 2.85) {$m=4$};

  \node at (v00) {$\varnothing$};
  \draw ($(v10)+(-0.13,-0.13)$) rectangle ($(v10)+(0.13,0.13)$);

  \draw ($(v20)+(-0.26,-0.13)$) rectangle ($(v20)+(0.26,0.13)$);
  \draw ($(v20)+(0,-0.13)$) -- ($(v20)+(0,0.13)$);

  \draw ($(v21)+(-0.13,-0.26)$) rectangle ($(v21)+(0.13,0.26)$);
  \draw ($(v21)+(-0.13,0)$) -- ($(v21)+(0.13,0)$);

  \draw ($(v30)+(-0.39,-0.13)$) rectangle ($(v30)+(0.39,0.13)$);
  \draw ($(v30)+(-0.13,-0.13)$) -- ($(v30)+(-0.13,0.13)$);
  \draw ($(v30)+(0.13,-0.13)$) -- ($(v30)+(0.13,0.13)$);

  \draw ($(v31)+(-0.26,0)$) rectangle ($(v31)+(0.26,0.26)$);
  \draw ($(v31)+(0,0)$) -- ($(v31)+(0,0.26)$);
  \draw ($(v31)+(-0.26,-0.26)$) rectangle ($(v31)+(0,0)$);

  \draw ($(v32)+(-0.13,-0.39)$) rectangle ($(v32)+(0.13,0.39)$);
  \draw ($(v32)+(-0.13,-0.13)$) -- ($(v32)+(0.13,-0.13)$);
  \draw ($(v32)+(-0.13,0.13)$) -- ($(v32)+(0.13,0.13)$);

  \draw ($(v40)+(-0.52,-0.13)$) rectangle ($(v40)+(0.52,0.13)$);
  \draw ($(v40)+(-0.26,-0.13)$) -- ($(v40)+(-0.26,0.13)$);
  \draw ($(v40)+(0,-0.13)$) -- ($(v40)+(0,0.13)$);
  \draw ($(v40)+(0.26,-0.13)$) -- ($(v40)+(0.26,0.13)$);

  \draw ($(v41)+(-0.39,0)$) rectangle ($(v41)+(0.39,0.26)$);
  \draw ($(v41)+(-0.13,0)$) -- ($(v41)+(-0.13,0.26)$);
  \draw ($(v41)+(0.13,0)$) -- ($(v41)+(0.13,0.26)$);
  \draw ($(v41)+(-0.39,-0.26)$) rectangle ($(v41)+(-0.13,0)$);

  \draw ($(v42)+(-0.26,-0.26)$) rectangle ($(v42)+(0.26,0.26)$);
  \draw ($(v42)+(0,-0.26)$) -- ($(v42)+(0,0.26)$);
  \draw ($(v42)+(-0.26,0)$) -- ($(v42)+(0.26,0)$);

  \draw ($(v43)+(-0.26,0.13)$) rectangle ($(v43)+(0.26,0.39)$);
  \draw ($(v43)+(0,0.13)$) -- ($(v43)+(0,0.39)$);
  \draw ($(v43)+(-0.26,-0.13)$) rectangle ($(v43)+(0,0.13)$);
  \draw ($(v43)+(-0.26,-0.39)$) rectangle ($(v43)+(0,-0.13)$);

  \draw ($(v44)+(-0.13,-0.52)$) rectangle ($(v44)+(0.13,0.52)$);
  \draw ($(v44)+(-0.13,-0.26)$) -- ($(v44)+(0.13,-0.26)$);
  \draw ($(v44)+(-0.13,0)$) -- ($(v44)+(0.13,0)$);
  \draw ($(v44)+(-0.13,0.26)$) -- ($(v44)+(0.13,0.26)$);

  \draw[bratteliarrow] (v00) -- (v10);
  \draw[bratteliarrow] (v10) -- (v20);
  \draw[bratteliarrow] (v10) -- (v21);
  \draw[bratteliarrow] (v20) -- (v30);
  \draw[bratteliarrow] (v20) -- (v31);
  \draw[bratteliarrow] (v21) -- (v31);
  \draw[bratteliarrow] (v21) -- (v32);
  \draw[bratteliarrow] (v30) -- (v40);
  \draw[bratteliarrow] (v30) -- (v41);
  \draw[bratteliarrow] (v31) -- (v41);
  \draw[bratteliarrow] (v31) -- (v42);
  \draw[bratteliarrow] (v31) -- (v43);
  \draw[bratteliarrow] (v32) -- (v43);
  \draw[bratteliarrow] (v32) -- (v44);
\end{tikzpicture} \caption{ Young lattice up to level $m=4$. The vertices at level $m$ are
Young diagrams $\lambda\vdash m$, and an edge $\nu\nearrow\lambda$
means that $\lambda$ is obtained from $\nu$ by adding one box. In
the symmetry-reduced $\mathsf{QC\text{-}QC}$ dual problem, these
edges specify the adjacent-level causal recursion between the variables
$s^{(m-1)}_{\nu}$ and $s^{(m)}_{\lambda}$. }
\label{fig:young_lattice_qcqc} 
\end{figure}
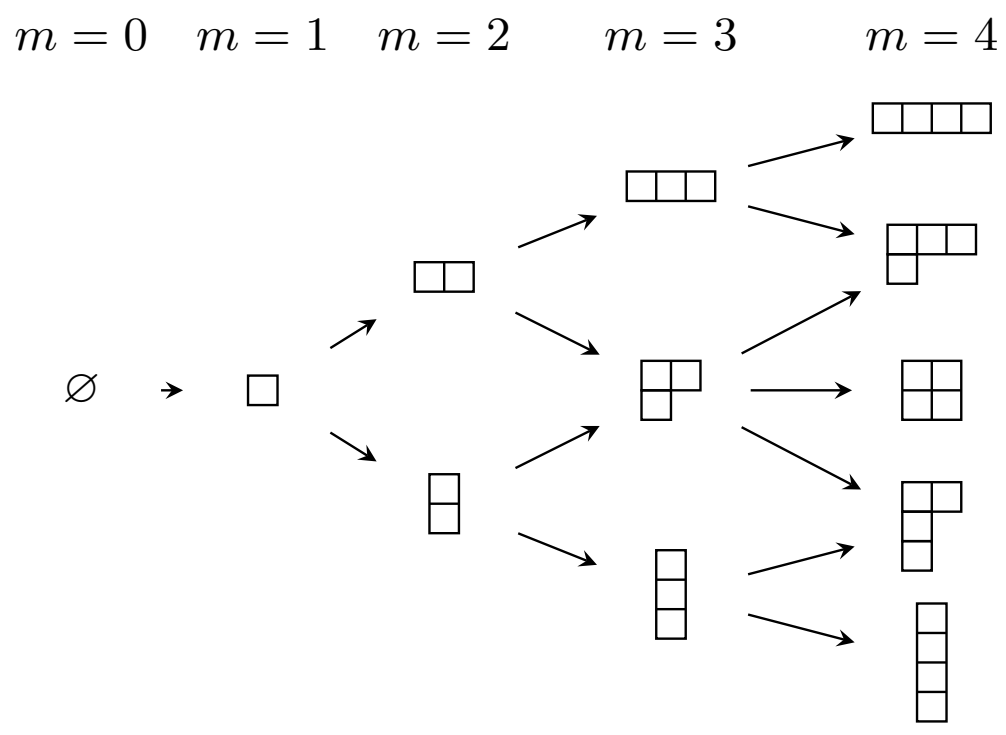

Thus, for $\mathsf{Bd}\in\{\mathsf{SLD},\mathsf{HN},\mathsf{NH}\}$,
the reduced dual problem for the $\mathsf{QC\text{-}QC}$ strategy
is 
\begin{equation}
\begin{aligned}
\max\quad&-\Tr h-d^N\mu\\
\mathrm{s.t.}\quad&B^{\mathsf{Bd}}_\lambda\succeq0,\\
&s^{(m-1)}_\nu\otimes I_{T^m_{\rm in}}\\
&\quad-\sum_{\lambda:\nu\nearrow\lambda}
\Phi^{(m)}_{\lambda\to\nu}(s^{(m)}_\lambda)\succeq0,\\
&d^N\mu I_{T^1_{\rm in}}\\
&\quad-\sum_{\lambda\in\mathbb Y_1^{(d^2)}}
\Phi^{(1)}_{\lambda\to\varnothing}(s^{(1)}_\lambda)\succeq0.
\end{aligned}
\label{eq:QCQC_reduced_psd_dual}
\end{equation}
The constraints run over $\lambda\in\mathbb Y_N^{(d^2)}$, $m=2,\ldots,N$, and $\nu\in\mathbb Y_{m-1}^{(d^2)}$, as appropriate. Here the top-level strategy-dual variables are $s^{(N)}_{\lambda}$. The adjacency structure of this recursion is organized by the Young lattice in Fig.~\ref{fig:young_lattice_qcqc}.

For the TCR bound, replace the block positivity constraints by Eq.~\eqref{eq:tcr_measurement_reduced_symred}. Group averaging, the canonical-chain reconstruction, and the Schur transform preserve feasibility and the objective value in both directions. Hence the reduced and original QC-QC dual problems have the same optimal value.

For fixed $q=d^2$, the number of Young diagrams at level $m$ is $O(m^{q-1})$, and the Weyl dimension formula bounds $\dim\mathcal U_\lambda^{(m)}$ by $O(m^{q(q-1)/2})$. Thus the total number of scalar entries in the history blocks is at most $O(N^{q^2})$. The precision-bound blocks also have polynomial size for fixed $d$ and $t$.

Parallel and general ICO strategies also admit full permutation reduction. Fixed-order strategies prescribe a chronological query sequence and therefore lack this symmetry. For causal-superposition strategies, permutation symmetry eliminates the need to impose $N!$ separate fixed-order constraints, while sequential constraints within each order still prevent a full polynomial block reduction. Table~\ref{tab:dual_complexity_scaling} compares the resulting sizes. Details for all strategy classes are collected in Appendix~\ref{app:schur-weyl}.

\begin{table}[t]
\centering
\caption{Growth of the dominant dual-variable size with the number of queries \(N\), for fixed \(d\) and \(t\).}
\label{tab:dual_complexity_scaling}
\renewcommand{\arraystretch}{1.18}
\begin{tabular}{@{}lccccc@{}}
\toprule
\(\text{status}\backslash \mathsf{Strat}\)
&
\(\mathsf{Para}\)
&
\(\mathsf{FO}\)
&
\(\mathsf{Sup}\)
&
\(\mathsf{QC\text{-}QC}\)
&
\(\mathsf{Gen}\)
\\
\midrule
\text{original}
&
\(\mathcal O(e^{cN})\)
&
\(\mathcal O(e^{cN})\)
&
\(\mathcal O(N!e^{cN})\)
&
\(\mathcal O(e^{cN})\)
&
\(\mathcal O(e^{cN})\)
\\
\text{reduced}
&
\(\mathcal O(N^c)\)
&
--
&
\(\mathcal O(e^{cN})\)
&
\(\mathcal O(N^c)\)
&
\(\mathcal O(N^c)\)
\\
\bottomrule
\end{tabular}
\end{table}

\subsection{Computable TCR bounds and physical recovery}

The TCR program retains a separable-cone constraint after symmetry reduction. We obtain a computable lower bound using the Doherty--Parrilo--Spedalieri (DPS) symmetric-extension hierarchy~\cite{doherty2004CompleteFamilySeparability} and an achievable upper bound using the AP inner approximation~\cite{ohst2024CertifyingQuantumSeparability}:
\begin{equation}
S(P_{\mathsf{DPS}_k}^{\mathsf{Strat}})
\le
S(P_{\mathsf{TCR}}^{\mathsf{Strat}})
\le
S(P_{\mathsf{AP}}^{\mathsf{Strat}}).
\label{eq:tcr_sandwich}
\end{equation}
At the first DPS level, the present construction reproduces the NH relaxation. An AP feasible point yields an explicit separable decomposition and hence a recoverable physical estimator. The Schur--Weyl blocks of Eq.~\eqref{eq:schur_weyl} apply to both approximations, allowing the two sides of Eq.~\eqref{eq:tcr_sandwich} to be evaluated in the reduced representation.

Suppose an AP iteration returns the feasible solution 
\begin{equation}
X_{{\rm AP}}=\sum^{M_{A}}_{\alpha=1}\sigma_{\alpha}\otimes\widetilde{\tau}_{\alpha},\label{eq:AP_recovery_X}
\end{equation}
where $\sigma_{\alpha}\succeq0$ are real symmetric operators and
$\widetilde{\tau}_{\alpha}\succeq0$. If the reduction based on Schur--Weyl
duality is used, first embed $\widetilde{\tau}_{\alpha}$ back into
$\mathcal{H}_{{\rm sys}}$.

Take a real rank-one decomposition $\sigma_{\alpha}=\sum_{\beta}|v_{\alpha\beta}\rangle\langle v_{\alpha\beta}|$,
and define $\mathcal{K}_{{\rm AP}}=\{(\alpha,\beta):\langle0|v_{\alpha\beta}\rangle\neq0\}$.
Following the recovery construction in Sec.~\ref{sec:tcr-recovery},
define for $(\alpha,\beta)\in\mathcal{K}_{{\rm AP}}$ 
\begin{align}
M_{\alpha\beta} & =\langle0|v_{\alpha\beta}\rangle^{2}\widetilde{\tau}_{\alpha},\label{eq:AP_recovery_outcome}\\
\hat{\theta}^{i}(\alpha,\beta) & =\theta^{i}+\frac{\langle i|v_{\alpha\beta}\rangle}{\langle0|v_{\alpha\beta}\rangle}.\label{eq:AP_recovery_estimator}
\end{align}
Because $X_{{\rm AP}}$ is feasible for the TCR-bound primal problem,
the recovery construction in Sec.~\ref{sec:tcr-recovery} guarantees that $\{M_{\alpha\beta},\hat{\theta}(\alpha,\beta)\}_{(\alpha,\beta)\in\mathcal{K}_{{\rm AP}}}$
is a valid locally unbiased estimation scheme for the chosen strategy,
and that 
\begin{equation}
\sum_{(\alpha,\beta)\in\mathcal{K}_{{\rm AP}}}M_{\alpha\beta}=X^{0,0}_{{\rm AP}}\in\mathcal{W}_{\mathsf{Strat}}.\label{eq:AP_recovery_normalization}
\end{equation}

Let the WMSE of this scheme be $C_{{\rm rec}}$. Applying again the
objective comparison in the recovery construction in Sec.~\ref{sec:tcr-recovery}, we obtain
\begin{equation}
S(P^{\mathsf{Strat}}_{\mathsf{TCR}})\leq C_{{\rm rec}}\leq\operatorname{Tr}[(G\otimes W_{\theta})X_{{\rm AP}}].\label{eq:AP_recovery_value_chain}
\end{equation}
The second inequality comes from discarding the nonnegative objective
contributions with $\langle0|v_{\alpha\beta}\rangle=0$. Equality
holds if all such contributions vanish. Thus an AP feasible point
gives both an upper bound on the TCR bound and an explicit process
POVM and classical estimator for the AP deterministic process.

\section{Numerical results}
\label{sec:numerics}

In this section, we numerically compare precision bounds across causal strategy classes for four noisy qubit-channel models using up to three queries. We examine the separation between different precision bounds, the dependence of the strategy hierarchy on the channel model, and the finite-resource precision advantage of QC-QCs over fixed-order strategies.

\subsection{Channel models}

In all numerical calculations, we use parameterized qubit channels
$\mathcal{E}_{\boldsymbol{\theta}}(\rho)=\sum_{a}K_{a}(\boldsymbol{\theta})\rho K^{\dagger}_{a}(\boldsymbol{\theta})$,
and compare $S(P^{\mathsf{Strat}}_{\mathsf{Bd}})$.

For the amplitude-damping (AD)-unitary channel, the parameters are
encoded by a three-axis rotation $U_{\boldsymbol{\omega}}=\exp\!\left[-\frac{i}{2}(\omega_{1}\sigma_{x}+\omega_{2}\sigma_{y}+\omega_{3}\sigma_{z})\right]$,
followed by amplitude damping: 
\begin{equation}
K_{0}=\begin{pmatrix}1 & 0\\
0 & \sqrt{1-\gamma}
\end{pmatrix}U_{\boldsymbol{\omega}},K_{1}=\begin{pmatrix}0 & \sqrt{\gamma}\\
0 & 0
\end{pmatrix}U_{\boldsymbol{\omega}}.
\end{equation}
Here $\boldsymbol{\omega}=(\omega_{1},\omega_{2},\omega_{3})$, and
$\gamma=0$ gives the purely unitary limit. 

The bit-flip (BF)-sandwich channel applies bit-flip noise both before
and after phase encoding: 
\begin{equation}
\begin{gathered}
U_\theta=e^{-i\theta\sigma_z/2},\\
B_0=\sqrt{1-p}\,\mathds{1},\quad B_1=\sqrt p\,\sigma_x,\\
K_{ab}=B_aU_\theta B_b,\quad a,b\in\{0,1\}.
\end{gathered}
\end{equation}
The estimated parameters are $(p,\theta)$.

The Z-rotation depolarizing channel is 
\begin{equation}
\begin{gathered}
U_\theta=e^{-i\theta\sigma_z/2},\quad K_0=\sqrt{1-\frac{3p}{4}}\,U_\theta,\\
K_j=\sqrt{\frac p4}\,\sigma_jU_\theta,\quad j=x,y,z.
\end{gathered}
\end{equation}
The estimated parameters are again $(p,\theta)$.

The AD-dephasing channel consists of a rotation about the $y$ axis,
amplitude damping, and a fixed dephasing background: 
\begin{equation}
\begin{gathered}U_{\Omega}=e^{-i\Omega\sigma_{y}/2},\;A_{0}=\begin{pmatrix}1 & 0\\
0 & \sqrt{1-p_{1}}
\end{pmatrix},\;A_{1}=\begin{pmatrix}0 & \sqrt{p_{1}}\\
0 & 0
\end{pmatrix},\\
K_0=\sqrt{1-q}\,A_0U_\Omega,\quad K_1=\sqrt{1-q}\,A_1U_\Omega,\\
K_2=\sqrt q\,\sigma_zA_0U_\Omega,\quad K_3=-\sqrt q\,A_1U_\Omega,\\
q=\frac14.
\end{gathered}
\end{equation}
The estimated parameters are $(p_{1},\Omega)$. The calculations below
use $N=1,2,3$, with the reduction based on Schur--Weyl duality above
whenever available. The four channel models are the AD-unitary,
BF-sandwich, Z-rotation depolarizing, and AD-dephasing channels.

\begin{figure}[tbp]
\centering
\includegraphics[width=\columnwidth]{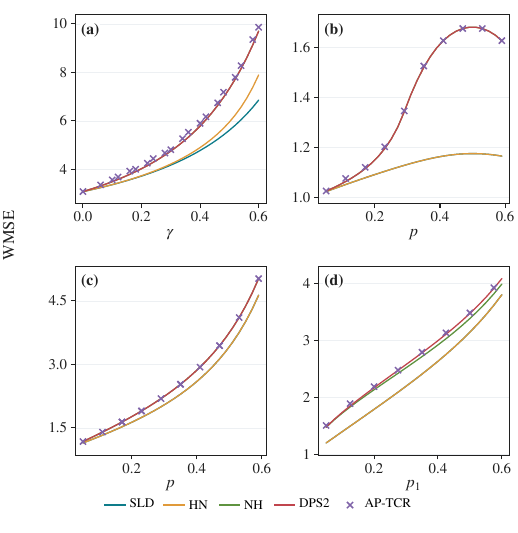}
\caption{All computed precision bounds for \(N=1\). Panels (a)--(d) show
the AD-unitary, BF-sandwich, Z-rotation depolarizing, and AD-dephasing
channels, respectively. The five strategy classes coincide for
one query. Colors identify SLD, HN, NH, DPS2, and the achievable AP-TCR
crosses.}
\label{fig:N1-bound-hierarchy-four-channels}
\end{figure}

Figure~\ref{fig:N1-bound-hierarchy-four-channels} gives the
one-query baseline for the four models. The bounds are tightly grouped
at weak noise and separate as the noise or control parameter increases.
The AP-TCR crosses follow the DPS level 2 (DPS2) curve in all four panels.

\subsection{Two-query precision advantage}

\begin{figure}[tbp]
\centering
\includegraphics[width=\columnwidth]{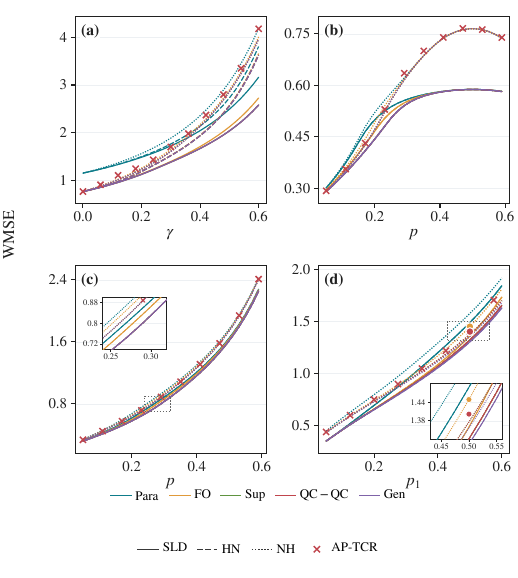}
\caption{All computed precision bounds for $N=2$ across the four channel
models. Panels (a)--(d) follow the channel order of
Fig.~\ref{fig:N1-bound-hierarchy-four-channels}. Colors identify the causal
strategy and line styles identify SLD, HN, and NH. The Sup and QC-QC
curves coincide within numerical precision for \(N=2\). AP-TCR samples for the QC-QC strategy are
shown as crosses. Dashed boxes mark the magnified regions. The FO--NH
value and the QC-QC AP-TCR value at $p_1=0.5$ are used in
Eq.~\eqref{eq:numerical_ico_advantage}.}
\label{fig:N2-strategy-hierarchy-four-channels-prl-cn}
\end{figure}

Figure~\ref{fig:N2-strategy-hierarchy-four-channels-prl-cn}(a) shows that
the Holevo bounds of the fixed-order and higher strategy classes agree within
numerical precision in the noiseless limit of the AD-unitary
channel. Fixed order already improves on parallel queries in this
limit. For the AD-dephasing channel in panel (d), the rotation angle
and amplitude-damping probability are estimated jointly with fixed
background dephasing. At \(p_1=0.5\) and \(q=1/4\), the AP upper bound
for QC-QC and the NH lower bound for fixed order give
\begin{equation}
S(P_{\mathsf{AP}}^{\QCQC}) \simeq 1.40182
< 1.44914 \simeq S(P_{\mathsf{NH}}^{\mathsf{FO}}).
\label{eq:numerical_ico_advantage}
\end{equation}
The inequality certifies that a physical QC-QC protocol achieves a
lower WMSE than any fixed-order protocol at that operating point.
\subsection{Separation between precision bounds}

\begin{figure}[tbp]
\centering
\includegraphics[width=\columnwidth]{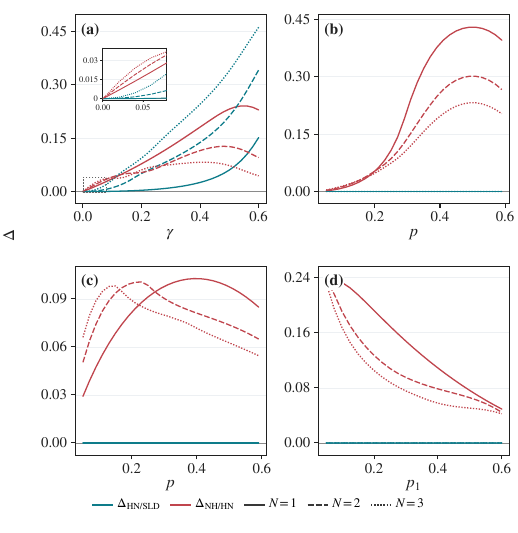}
\caption{Normalized gaps between precision bounds. Panels (a)--(d) follow
the channel order of Fig.~\ref{fig:N1-bound-hierarchy-four-channels}.
Line styles distinguish the common one-query strategy from fixed order at
\(N=2\) and \(N=3\).}
\label{fig:bound-gap-cross-N}
\end{figure}

To compare bound separations at different query
counts, define the relative gaps at fixed \(N\) and
\(\mathsf{Strat}\) as
\begin{equation}
\begin{aligned}
\Delta^{(N,\mathsf{Strat})}_{\mathsf{HN/SLD}}
  &:= \frac{S(P^{\mathsf{Strat}}_{\mathsf{HN}})-S(P^{\mathsf{Strat}}_{\mathsf{SLD}})}
            {S(P^{\mathsf{Strat}}_{\mathsf{SLD}})},\\
\Delta^{(N,\mathsf{Strat})}_{\mathsf{NH/HN}}
  &:= \frac{S(P^{\mathsf{Strat}}_{\mathsf{NH}})-S(P^{\mathsf{Strat}}_{\mathsf{HN}})}
            {S(P^{\mathsf{Strat}}_{\mathsf{HN}})} .
\end{aligned}
\label{eq:normalized-bound-gaps}
\end{equation}
The first gap quantifies the increase in the lower bound when
measurement incompatibility enters the optimization. The second
quantifies the additional tightening from the finite-measurement
constraints of the NH relaxation.
Figure~\ref{fig:bound-gap-cross-N} compares the common one-query
values with fixed-order results at \(N=2,3\).

For the AD-unitary channel, \(\Delta_{\mathsf{HN/SLD}}\) grows with
the damping strength \(\gamma\), with a larger increase at \(N=2\)
and \(N=3\) than at \(N=1\) over the scanned range. The gap
\(\Delta_{\mathsf{NH/HN}}\) is also appreciable.

For the other three channels, \(\Delta_{\mathsf{HN/SLD}}\) is close
to zero within numerical precision, and \(\Delta_{\mathsf{NH/HN}}\)
accounts for most of the separation. The latter gap peaks at
an intermediate flip probability for the BF-sandwich channel, exhibits
a weaker nonmonotonic dependence on \(p\) for the Z-rotation
depolarizing channel, and decreases over most of the AD-dephasing
scan.

\subsection{Strategy hierarchy and ICO advantage}

\begin{figure}[tbp]
\centering
\includegraphics[width=\columnwidth]{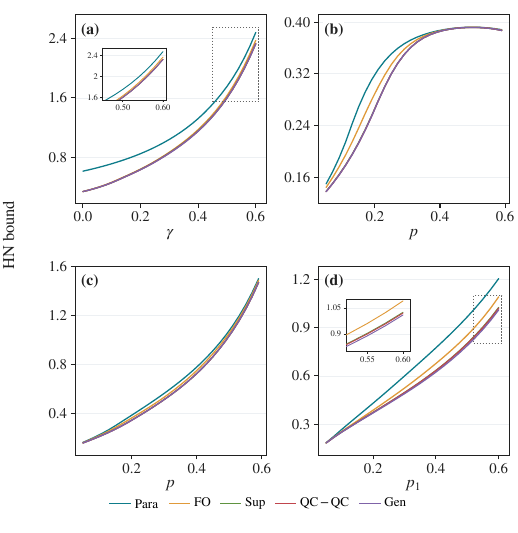}
\caption{Holevo bounds \(S(P^{\mathsf{Strat}}_{\mathsf{HN}})\) at
\(N=3\) for the five strategy classes. Panels (a)--(d) follow the channel
order of Fig.~\ref{fig:N1-bound-hierarchy-four-channels}.}
\label{fig:N3-strategy-hierarchy-four-channels}
\end{figure}

Figure~\ref{fig:N3-strategy-hierarchy-four-channels} compares the five
strategy classes at \(N=3\) using the Holevo bound. To resolve the
nearby curves, let \(\mathsf S_r\), \(r=0,\ldots,4\), denote the
strategies in the order of Eq.~\eqref{eq:strategy_hierarchy}. The
relative improvement from each extension is
\begin{equation}
\eta^{(N,\mathsf{HN})}_r
:= \frac{S(P^{\mathsf S_r}_{\mathsf{HN}})
             -S(P^{\mathsf S_{r+1}}_{\mathsf{HN}})}
            {S(P^{\mathsf S_r}_{\mathsf{HN}})},
\qquad r=0,\ldots,3 .
\label{eq:adjacent-strategy-gain}
\end{equation}
For \(N=2\), \(\mathsf{Sup}=\QCQC\), so
\(\eta^{(2,\mathsf{HN})}_2=0\).

\begin{figure}[tbp]
\centering
\includegraphics[width=\columnwidth]{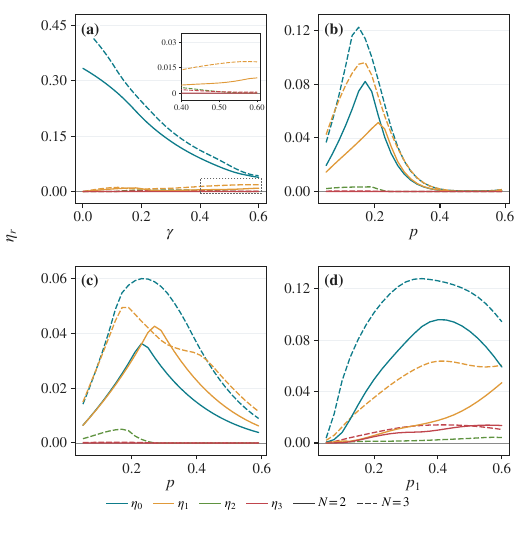}
\caption{Relative gains \(\eta^{(N,\mathsf{HN})}_r\) between adjacent
strategy classes. Panels (a)--(d) follow the channel order of
Fig.~\ref{fig:N1-bound-hierarchy-four-channels}. Colors represent the
four extensions in Eq.~\eqref{eq:adjacent-strategy-gain} from parallel
to general ICO; straight segments join the stored data points. Solid and
dashed curves correspond to \(N=2\) and \(N=3\). The zero \(\eta^{(2,\mathsf{HN})}_2\) curve is omitted.}
\label{fig:strategy-gap-cross-N}
\end{figure}

Figure~\ref{fig:strategy-gap-cross-N} shows the adjacent-strategy gains.
At the unitary point \(\gamma=0\) of the AD-unitary model,
\(\eta_1,\eta_2,\eta_3\) vanish within numerical precision for both
query counts. The fixed-order Holevo bound there already reaches
the value of the general ICO bound. In the noisy part of the
scans, the largest gains usually occur at the
\(\mathsf{Para}\to\mathsf{FO}\) and
\(\mathsf{FO}\to\mathsf{Sup}\) steps. Further extensions yield smaller
reductions of the Holevo bound, whose sizes depend on the channel
and operating point.

\section{Conclusion}
\label{sec:conclusion}

In this work, we have developed a symmetry-reduced framework for multiparameter quantum channel estimation with QC-QCs. By exploiting the permutation symmetry of repeated queries of an identical channel and reformulating the recursive causal constraints in Schur--Weyl blocks, we obtain an exact reduction of the size of the QC-QC dual optimization problem from exponential to polynomial in the number of queries. Combined with established conic formulations of multiparameter precision bounds, this framework enables systematic comparisons across different strategy classes in both the asymptotic and finite-resource regimes.

Our numerical results demonstrate that the precision advantage of ICO depends on the channel model and the precision bound under consideration. In particular, with only two channel queries, we obtain a physically achievable QC-QC WMSE via the AP-TCR formulation that lies below the fixed-order NH bound, thereby certifying a strict finite-resource precision advantage. We believe these results will be useful for practical multiparameter estimation of quantum channels.

\section*{Data availability}
The numerical data supporting the findings of this article can be reproduced using the publicly available source code and simulation parameters~\cite{multiparameterMetrologyCode}.

\section*{Acknowledgments}
 This work is supported by the National Natural Science Foundation of China (Grant No. 12075323), the Natural Science Foundation of Guangdong Province of China (Grant No. 2025A1515011440), and the Innovation Program for Quantum Science and Technology (Grant No. 2021ZD0300702).

\appendix

\section{Equivalence between the two-level optimization and the single-level
conic program}

\label{app:single-double}

This section proves that the two-level optimization obtained by using
the final state optimization of Hayashi~\cite{hayashi2023TightCramerRaoType}
as the inner problem and optimizing the strategy in the outer problem
is equivalent to the single-level conic program in the main text.

\begin{proof}[Proof of Theorem~\ref{thm:two_level_single_level_equivalence}]
We first prove $S(P^{\mathsf{Strat}}_{\mathsf{Bd}})\leq K(P^{\mathsf{Strat}}_{\mathsf{Bd}})$.
Take any feasible point $(\widetilde{P},T)$ of the two-level problem,
and define $\Lambda_{\widetilde{P}}(A):=\operatorname{Tr}_{F}\!\left[\widetilde{P}(I_{{\rm sys}}\otimes A^{\mathsf{T}})\right].$
Since $\widetilde{P}\succeq0$, $\Lambda_{\widetilde{P}}$ is completely
positive, and $\Lambda_{\widetilde{P}}(I_{F})=\operatorname{Tr}_{F}\widetilde{P},\;\Lambda^{*}_{\widetilde{P}}(B)=\left\{ \operatorname{Tr}_{{\rm sys}}[(B\otimes I_{F})\widetilde{P}]\right\} ^{\mathsf{T}}.$
Set 
\begin{equation}
X=(\operatorname{id}_{\mathcal{R}}\otimes\Lambda_{\widetilde{P}})(T^{\mathsf{T}}).\label{eq:equivalence_forward_construction}
\end{equation}
Then, for any $A\in\mathcal{L}(\mathcal{R})$ and $B\in\mathcal{L}(\mathcal{H}_{{\rm sys}})$,
\begin{equation}
\operatorname{Tr}[(A\otimes B)X]=\operatorname{Tr}\!\left[T\left(A^{\mathsf{T}}\otimes\operatorname{Tr}_{{\rm sys}}[(B\otimes I_{F})\widetilde{P}]\right)\right].\label{eq:equivalence_forward_pairing}
\end{equation}
Thus $X^{0,0}=\operatorname{Tr}_{F}\widetilde{P}\in\mathcal{W}_{\mathsf{Strat}},\;\operatorname{Tr}X^{0,0}=d^N.$
Introduce ${\displaystyle K_{i}:=\frac{1}{2}\left(|0\rangle\langle i|+|i\rangle\langle0|\right)}$,
$i=0,\ldots,t$. Substituting \((A,B)=(K_{i},\partial_{j}W_{\theta})\)
and \((A,B)=(G,W_{\theta})\) into Eq.~\eqref{eq:equivalence_forward_pairing}
shows that the local-unbiasedness constraints and the objective value
are preserved. The transpose and the completely positive map on the
physical system preserve the corresponding cone constraints for SLD
and NH. The additional HN constraint follows from $R^{\mathsf{T}}_{kl}=-R_{kl}$
and the same trace identity. Hence $X$ is a feasible point of the
single-level problem with the same value, and therefore 
\begin{equation}
S(P^{\mathsf{Strat}}_{\mathsf{Bd}})\leq K(P^{\mathsf{Strat}}_{\mathsf{Bd}}).\label{eq:equivalence_forward_bound}
\end{equation}

We next prove the reverse inequality. Take any feasible point $X$
of the single-level problem, and choose a strictly positive normalized
strategy point $P_{0}\in\mathcal{W}_{\mathsf{Strat}}$. For $0<\varepsilon<1$,
define $L_{\varepsilon}=\sqrt{1-\varepsilon}|0\rangle\langle0|+\frac{1}{\sqrt{1-\varepsilon}}\sum^{t}_{i=1}|i\rangle\langle i|,$
\begin{equation}
X_{\varepsilon}=(L_{\varepsilon}\otimes I_{{\rm sys}})X(L_{\varepsilon}\otimes I_{{\rm sys}})+\varepsilon|0\rangle\langle0|\otimes P_{0}.\label{eq:equivalence_regularized_X}
\end{equation}
Then $X_{\varepsilon}$ remains feasible and $P_{\varepsilon}:=X^{0,0}_{\varepsilon}=(1-\varepsilon)X^{0,0}+\varepsilon P_{0}\succ0,\;X^{0,i}_{\varepsilon}=X^{0,i}.$
Since $G|0\rangle=0$, 
\begin{equation}
\operatorname{Tr}[(G\otimes W_{\theta})X_{\varepsilon}]=\frac{1}{1-\varepsilon}\operatorname{Tr}[(G\otimes W_{\theta})X].\label{eq:equivalence_regularized_objective}
\end{equation}

Take $\mathcal{H}_{F}\simeq\mathcal{H}_{{\rm sys}}$, and set $|\Psi_{\varepsilon}\rangle=(P^{1/2}_{\varepsilon}\otimes I_{F})\sum_{a}|a\rangle_{{\rm sys}}|a\rangle_{F},\;\widetilde{P}_{\varepsilon}=|\Psi_{\varepsilon}\rangle\langle\Psi_{\varepsilon}|.$
Then $\operatorname{Tr}_{F}\widetilde{P}_{\varepsilon}=P_{\varepsilon}$.
Define 
\begin{equation}
T_{\varepsilon}=\left[(I_{\mathcal{R}}\otimes P^{-1/2}_{\varepsilon})X_{\varepsilon}(I_{\mathcal{R}}\otimes P^{-1/2}_{\varepsilon})\right]^{\mathsf{T}}.\label{eq:equivalence_backward_T}
\end{equation}
Then $T^{0,0}_{\varepsilon}=I_{F}$, and the standard purification
gives 
\begin{equation}
\operatorname{Tr}\!\left[T_{\varepsilon}\left(A\otimes\operatorname{Tr}_{{\rm sys}}[(B\otimes I_{F})\widetilde{P}_{\varepsilon}]\right)\right]=\operatorname{Tr}\!\left[X_{\varepsilon}(A^{\mathsf{T}}\otimes B)\right].\label{eq:equivalence_pairing_identity}
\end{equation}
We now check the constraints of the two-level problem. Since $K^{\mathsf{T}}_{i}=K_{i}$,
taking $A=K_{i}$ and $B=\partial_{j}W_{\theta}$ in Eq.~\eqref{eq:equivalence_pairing_identity}
gives 
\begin{equation}
\operatorname{Tr}\!\left[T_{\varepsilon}(K_{i}\otimes\partial_{j}\rho_{\theta}(\widetilde{P}_{\varepsilon}))\right]=\operatorname{Tr}\!\left[X_{\varepsilon}(K_{i}\otimes\partial_{j}W_{\theta})\right]=\delta_{ij},
\end{equation}
which is exactly local unbiasedness for the final-state model $\rho_{\theta}(\widetilde{P}_{\varepsilon})$.
Since the weight matrix is real symmetric, $G^{\mathsf{T}}=G$. Taking
$A=G$ and $B=W_{\theta}$ gives 
\begin{equation}
\operatorname{Tr}\!\left[T_{\varepsilon}(G\otimes\rho_{\theta}(\widetilde{P}_{\varepsilon}))\right]=\operatorname{Tr}\!\left[X_{\varepsilon}(G\otimes W_{\theta})\right],
\end{equation}
so the value of the inner estimation problem is the same as the value
of the single-level variable $X_{\varepsilon}$. For the HN bound,
the additional real-block constraints are also preserved: using $R^{\mathsf{T}}_{kl}=-R_{kl}$
and $B=W_{\theta}$ gives the required zero constraint.

It remains to check that $T_{\varepsilon}$ satisfies the precision-bound
cone constraint in the two-level output-state estimation problem.
By Eq.~\eqref{eq:equivalence_backward_T}, $T_{\varepsilon}$ is
obtained by conjugating $X_{\varepsilon}$ with $I_{\mathcal{R}}\otimes P^{-1/2}_{\varepsilon}$,
then taking the full transpose and relabeling the second system as
$F$. Conjugation by an invertible operator and the transpose preserve
positivity. The same operations preserve the block conditions used
for the SLD, HN, and NH bounds. Therefore $T_{\varepsilon}$ satisfies
the precision-bound cone constraint for the chosen bound, and $(\widetilde{P}_{\varepsilon},T_{\varepsilon})$
is feasible for the two-level problem. Hence $K(P^{\mathsf{Strat}}_{\mathsf{Bd}})\leq\frac{1}{1-\varepsilon}\operatorname{Tr}[(G\otimes W_{\theta})X].$
Letting $\varepsilon\to0$ and then taking the infimum over all single-level
feasible points gives 
\begin{equation}
K(P^{\mathsf{Strat}}_{\mathsf{Bd}})\leq S(P^{\mathsf{Strat}}_{\mathsf{Bd}}).\label{eq:equivalence_backward_bound}
\end{equation}
Combining Eqs.~\eqref{eq:equivalence_forward_bound} and~\eqref{eq:equivalence_backward_bound}
proves the claim. 
\end{proof}

\section{Strategy dual cones and fixed-order projections}

\label{app:strategy-cones}

The primal strategy cones are defined in Sec.~\ref{sec:strategy-cones}. This appendix collects the dual characterizations and the fixed-order projection maps.

The dual strategy cone is defined by 
\begin{equation}
\mathcal{W}^{*}_{\mathsf{Strat}}:=\left\{ S=S^{\dagger}\ \middle|\ \Tr[SP]\ge0,\ \forall P\in\mathcal{W}_{\mathsf{Strat}}\right\} .\label{eq:strategy_dual_cone_definition}
\end{equation}

\paragraph*{Parallel strategies $(\mathsf{Para})$.}

The dual cone is 
\begin{equation}
\mathcal{W}^{*}_{\mathsf{Para}}=\left\{ S=S^{\dagger}\ \middle|\ \Tr_{T^{1}_{{\rm out}}\cdots T^{N}_{{\rm out}}}[S]\succeq0\right\} .\label{eq:dual_para}
\end{equation}

\paragraph*{Fixed-order strategies $(\mathsf{FO})$.}

Equivalently, the fixed-order cone can be described by a projection
condition for the linear subspace of quantum combs: 
\begin{equation}
\mathcal{W}^{(e)}_{\mathsf{FO}}=\left\{ P\succeq0\ \middle|\ \mathcal{Q}^{(e)}_{\mathsf{FO}}(P)=0\right\} ,\;\mathcal{Q}^{(e)}_{\mathsf{FO}}:=\mathcal{I}-\mathcal{P}^{(e)}_{\mathsf{FO}}.\label{eq:cone_fo_projection}
\end{equation}
Here the projection superoperator for the fixed order $e=(1,2,\ldots,N)$
is 
\begin{widetext}
\begin{equation}
\begin{aligned}\mathcal{P}^{(e)}_{\mathsf{FO}} & =\mathcal{D}_{T^{N}_{\mathrm{out}}}\left[\mathcal{I}-\mathcal{D}_{T^{N}_{\mathrm{in}}}\left(\mathcal{I}-\mathcal{D}_{T^{N-1}_{\mathrm{out}}}\left[\mathcal{I}-\mathcal{D}_{T^{N-1}_{\mathrm{in}}}\left(\cdots\mathcal{I}-\mathcal{D}_{T^{2}_{\mathrm{out}}}\left[\mathcal{I}-\mathcal{D}_{T^{2}_{\mathrm{in}}}\left(\mathcal{I}-\mathcal{D}_{T^{1}_{\mathrm{out}}}\right)\right]\cdots\right)\right]\right)\right]\\
 & =\mathcal{D}_{T^{N}_{\mathrm{out}}}-\mathcal{D}_{T^{N}_{\mathrm{in}}T^{N}_{\mathrm{out}}}+\mathcal{D}_{T^{N-1}_{\mathrm{out}}T^{N}_{\mathrm{in}}T^{N}_{\mathrm{out}}}-\mathcal{D}_{T^{N-1}_{\mathrm{in}}T^{N-1}_{\mathrm{out}}T^{N}_{\mathrm{in}}T^{N}_{\mathrm{out}}}+\cdots+\mathcal{D}_{T^{1}_{\mathrm{out}}T^{2}_{\mathrm{in}}T^{2}_{\mathrm{out}}\cdots T^{N}_{\mathrm{in}}T^{N}_{\mathrm{out}}}.
\end{aligned}
\label{eq:FO_projection}
\end{equation}
\end{widetext}
For any permutation $\pi\in\mathfrak{S}_{N}$, replacing the order
in Eq.~\eqref{eq:FO_projection} by $(\pi(1),\pi(2),\ldots,\pi(N))$
gives $\mathcal{P}^{(\pi)}_{\mathsf{FO}}$, $\mathcal{Q}^{(\pi)}_{\mathsf{FO}}$,
and $\mathcal{W}^{(\pi)}_{\mathsf{FO}}$.

The dual cone of a fixed-order cone also has two equivalent descriptions.
The projection form gives 
\begin{equation}
\left(\mathcal{W}^{(e)}_{\mathsf{FO}}\right)^{*}=\left\{ S=S^{\dagger}\ \middle|\ \exists Z,\ S-\mathcal{Q}^{(e)}_{\mathsf{FO}}(Z)\succeq0\right\} .\label{eq:dual_fo_abstract}
\end{equation}
The nested linear-matrix-inequality (LMI) form is 
\begin{equation}
\begin{aligned}
\left(\mathcal{W}^{(e)}_{\mathsf{FO}}\right)^{*}={}&\Bigl\{ S=S^{\dagger}\mid\exists\{\Lambda^{(n)}\}^{N}_{n=1}\ \text{with}\ \Lambda^{(N)}=S,\\
&\quad \Tr_{T^{n}_{{\rm out}}}[\Lambda^{(n)}]-\Lambda^{(n-1)}\otimes I_{T^{n}_{{\rm in}}}\succeq0\\
&\quad (n=2,\ldots,N),\\
&\quad \Tr_{T^{1}_{{\rm out}}}[\Lambda^{(1)}]\succeq0\Bigr\} .
\end{aligned}
\label{eq:dual_fo_explicit}
\end{equation}

\paragraph*{Causal-superposition strategies $(\mathsf{Sup})$.}

The dual cone is therefore the intersection of the fixed-order dual
cones: 
\begin{equation}
\mathcal{W}^{*}_{\mathsf{Sup}}=\bigcap_{\pi\in\mathfrak{S}_{N}}\left(\mathcal{W}^{(\pi)}_{\mathsf{FO}}\right)^{*}.\label{eq:dual_sup}
\end{equation}

\paragraph*{Quantum circuits with quantum control of causal order $(\mathsf{QC\text{-}QC})$
strategies.}

The dual cone corresponding to Eqs.~\eqref{eq:cone_qcqc_top} and~\eqref{eq:cone_qcqc_rec} is
\begin{equation}
\begin{aligned}
\mathcal{W}^{*}_{\mathsf{QC\text{-}QC}}={}&\Bigl\{ S=S^{\dagger}\mid\exists\,\{S_{\mathcal{K}}\}_{\varnothing\neq\mathcal{K}\subsetneq\mathcal{N}}\ \text{with}\ S_{\mathcal{N}}=S,\\
&\quad \Tr_{T^{k}_{\mathrm{out}}}\!\bigl[S_{\mathcal{K}\cup\{k\}}\bigr]-S_{\mathcal{K}}\otimes I_{T^{k}_{\mathrm{in}}}\succeq0,\\
&\quad \forall\,\varnothing\neq\mathcal{K}\subsetneq\mathcal{N},\ k\in\mathcal{N}\setminus\mathcal{K},\\
&\quad \Tr_{T^{k}_{\mathrm{out}}}\!\bigl[S_{\{k\}}\bigr]\succeq0,\ \forall\,k\in\mathcal{N}\Bigr\} .
\end{aligned}
\label{eq:dual_qcqc}
\end{equation}

\paragraph*{General indefinite-causal-order strategies $(\mathsf{Gen})$.}

The dual cone is 
\begin{equation}
\mathcal{W}^{*}_{\mathsf{Gen}}=\left\{ S=S^{\dagger}\ \middle|\ \exists Z,\ S-\mathcal{Q}_{\mathsf{Gen}}(Z)\succeq0\right\} .\label{eq:dual_gen}
\end{equation}

In the unified dual problem~\eqref{eq:unified_dual}, changing the
strategy class only changes the strategy-dual constraint. 

\section{Proof of strong duality: explicit Slater construction}

\label{app:slater_condition}

This section establishes the strong-duality statement in the main text
by explicitly constructing strictly feasible points.

\begin{theorem}[Strong duality] \label{thm:strong_duality} Assume
that the first derivatives of the Choi operator, $\partial_{1}W_{\theta},\ldots,\partial_{t}W_{\theta}$,
are linearly independent. Then, for $\mathsf{Bd}\in\{\mathsf{SLD},\mathsf{HN},\mathsf{NH},\mathsf{TCR}\}$
and $\mathsf{Strat}\in\{\mathsf{Para},\mathsf{FO},\mathsf{Sup},\mathsf{QC\text{-}QC},\mathsf{Gen}\}$,
the unified primal problem $P^{\mathsf{Strat}}_{\mathsf{Bd}}$
satisfies the primal Slater condition. Therefore, $P^{\mathsf{Strat}}_{\mathsf{Bd}}$
and $D^{\mathsf{Strat}}_{\mathsf{Bd}}$ have the same optimal value,
and the dual optimum is attained. \end{theorem}
\begin{proof}
Take $P_{{\rm mix}}=d^{-N}I_{\mathcal{H}_{{\rm sys}}}$, so $\operatorname{Tr}P_{{\rm mix}}=d^{N}$.

For the parallel strategy, $P_{{\rm mix}}=d^{-N}I_{{\rm in}}\otimes I_{{\rm out}}$,
so the normalization constraint holds. For fixed-order strategies,
in the standard order take $P^{(n)}=d^{-n}I_{\bigotimes^{n-1}_{r=1}(T^{r}_{{\rm in}}\otimes T^{r}_{{\rm out}})\otimes T^{n}_{{\rm in}}}$.
The operators $P^{(n)}$ are strictly positive and satisfy $\operatorname{Tr}_{T^{n+1}_{{\rm in}}}P^{(n+1)}=P^{(n)}\otimes I_{T^{n}_{{\rm out}}}$,
$\operatorname{Tr}_{T^{1}_{{\rm in}}}P^{(1)}=1$ and $P^{(N)}\otimes I_{T^{N}_{{\rm out}}}=P_{{\rm mix}}$.
The other fixed orders are obtained by permuting the ports.

For causal-superposition strategies, for each $\pi\in\mathfrak{S}_{N}$
take $Y_{\pi}=P_{{\rm mix}}/N!$. Each $Y_{\pi}$ is strictly positive,
satisfies the homogeneous causal constraints of the corresponding
fixed order, and $\sum_{\pi\in\mathfrak{S}_{N}}Y_{\pi}=P_{{\rm mix}}$.
For general process strategies, the identity is invariant under the
local replacement maps, so $\mathcal{Q}_{\mathsf{Gen}}(P_{{\rm mix}})=0$.

For $\mathsf{QC\text{-}QC}$ strategies, when $|\mathcal{K}_{n-1}|=n-1$
and $k_{n}\notin\mathcal{K}_{n-1}$ take 
\begin{equation}
W_{(\mathcal{K}_{n-1},k_{n})}=\frac{I_{\bigotimes_{r\in\mathcal{K}_{n-1}}(T^{r}_{{\rm in}}\otimes T^{r}_{{\rm out}})\otimes T^{k_{n}}_{{\rm in}}}}{n\binom{N}{n}d^{n}}.\label{eq:slater_qcqc_variables}
\end{equation}
The node variables $W_{(\mathcal{K}_{n-1},k_{n})}$ are strictly positive.
Let $c_{n}=[n\binom{N}{n}d^{n}]^{-1}$. The initial level satisfies
$\sum^{N}_{k_{1}=1}\operatorname{Tr}_{T^{k_{1}}_{{\rm in}}}W_{(\varnothing,k_{1})}=1$.
The coefficients on the two sides of the recursive constraint are
$(N-n)dc_{n+1}$ and $nc_{n}$. The two coefficients are equal because
$(n+1)\binom{N}{n+1}=(N-n)\binom{N}{n}$. The final level satisfies
$\sum^{N}_{k_{N}=1}W_{(\{1,\ldots,N\}\setminus\{k_{N}\},k_{N})}\otimes I_{T^{k_{N}}_{{\rm out}}}=P_{{\rm mix}}$.
Thus all five strategy classes admit strictly positive lifted variables
satisfying all affine constraints.

We next treat SLD, HN, and NH together. Since the derivative operators
are linearly independent, there exist Hermitian operators $Y_{1},\ldots,Y_{t}$
satisfying 
\begin{equation}
\operatorname{Tr}[\partial_{k}W_{\theta}Y_{j}]=\delta_{kj}\ \text{for}\ j,k=1,\ldots,t.\label{eq:slater_derivative_dual_basis}
\end{equation}
Define 
\begin{equation}
X_{\eta}=\begin{pmatrix}P_{{\rm mix}} & Y_{1} & \cdots & Y_{t}\\
Y_{1} & \eta I_{{\rm sys}} & \cdots & 0\\
\vdots & \vdots & \ddots & \vdots\\
Y_{t} & 0 & \cdots & \eta I_{{\rm sys}}
\end{pmatrix}.\label{eq:slater_sdp_point}
\end{equation}
Let $\mathbf{Y}=(Y_{1},\ldots,Y_{t})$. When $\eta>\|\mathbf{Y}^{\dagger}P^{-1}_{{\rm mix}}\mathbf{Y}\|_{\infty}$
the Schur complement gives $X_{\eta}\succ0$. Moreover, $(X_{\eta})^{0,0}=P_{{\rm mix}}$
and $\operatorname{Tr}[\partial_{k}W_{\theta}(X_{\eta})^{0,j}]=\delta_{kj}$.

The operator $X_{\eta}$ satisfies the block-linear constraints required
by NH. Since the NH block-linear constraints are the strongest, $X_{\eta}$
also belongs to $\operatorname{ri}(\mathcal{S}_{\mathsf{NH}})$, $\operatorname{ri}(\mathcal{S}_{\mathsf{HN}})$
and $\operatorname{ri}(\mathcal{S}_{\mathsf{SLD}})$. Hence the three
primal problems satisfy Slater's condition.

Finally consider the TCR bound. Take a full-rank informationally complete
positive-operator-valued measure (POVM) $\{E_{x}\}_{x\in\mathcal{X}}$,
satisfying $E_{x}\succ0$, $\sum_{x}E_{x}=I_{{\rm sys}}$, and spanning
the real vector space of Hermitian operators on the system. Define
\begin{equation}
F_{jx}=d^{-N}\operatorname{Tr}[\partial_{j}W_{\theta}E_{x}].\label{eq:slater_response_matrix}
\end{equation}
By informational completeness and real linear independence of the
derivative operators, $F$ has full row rank. Hence there exist real
numbers $c^{i}_{x}$ satisfying 
\begin{equation}
\sum_{x\in\mathcal{X}}c^{i}_{x}F_{jx}=\delta_{ij}.\label{eq:slater_response_inverse}
\end{equation}

Choose any $\gamma>1$, let $|\psi^{(\gamma)}_{x}\rangle=|0\rangle+\gamma\sum^{t}_{i=1}c^{i}_{x}|i\rangle$,
and define 
\begin{equation}
X^{(\gamma)}=d^{-N}\sum_{x\in\mathcal{X}}|\psi^{(\gamma)}_{x}\rangle\langle\psi^{(\gamma)}_{x}|\otimes E_{x}.\label{eq:slater_scaled_tcr_point}
\end{equation}
The operator $X^{(\gamma)}$ lies in $\mathcal{S}_{{\rm SEP}}(\mathcal{R}:\mathcal{H}_{{\rm sys}})\cap\mathcal{B}=\mathcal{S}_{\mathsf{TCR}}$,
and satisfies $(X^{(\gamma)})^{0,0}=P_{{\rm mix}}$ and $\operatorname{Tr}[\partial_{j}W_{\theta}(X^{(\gamma)})^{0,i}]=\gamma\delta_{ij}$.

Let $X_{{\rm id}}=I_{\mathcal{R}}\otimes P_{{\rm mix}}$. The product
identity lies in the interior of the separable cone, so $X_{{\rm id}}\in\operatorname{ri}(\mathcal{S}_{\mathsf{TCR}})$.
Define 
\begin{equation}
X^{*}=\gamma^{-1}X^{(\gamma)}+(1-\gamma^{-1})X_{{\rm id}}.\label{eq:slater_tcr_final_point}
\end{equation}
By convexity of the relative interior, $X^{*}\in\operatorname{ri}(\mathcal{S}_{\mathsf{TCR}})$.
At the same time, $(X^{*})^{0,0}=P_{{\rm mix}}$ and $\operatorname{Tr}[\partial_{j}W_{\theta}(X^{*})^{0,i}]=\delta_{ij}$.
Pairing $X^{*}$ with the strictly positive lifted strategy variables
constructed above gives a full Slater point for the TCR-bound primal
problem.

Thus the unified primal problems for all four precision bounds and
all five strategy classes satisfy the primal Slater condition. By
Slater's theorem for finite-dimensional closed conic programs, the
primal and dual problems have no duality gap, and the dual optimum
is attained. 
\end{proof}

\section{Symmetry reductions for the other strategy classes}

\label{app:schur-weyl}

This appendix uses the Schur-block notation, embeddings, and precision-bound constraints defined in Sec.~\ref{sec:symmetry}. Let $\mathsf U_{\rm in}$ denote the input-side Schur transform.

\subsection{Para strategy}

The causal constraint $\mu I_{{\rm sys}}-S\in\mathcal{W}^{*}_{\mathsf{Para}}$
in the unified dual is equivalent to 
\begin{equation}
\mu d^{N}I_{{\rm in}}-\operatorname{Tr}_{{\rm out}}S\succeq0.\label{eq:para_unreduced_causal_constraint_symred}
\end{equation}

For a permutation-invariant operator $Y$ on the input space, let
$[Y]^{{\rm in}}_{\nu}$ be its $\nu$-th Schur block under the $\mathsf{U}_{{\rm in}}$
decomposition. For $x\in\mathcal{L}(\mathcal{U}_{\lambda})$, let
$\iota_{\lambda}(x)$ be the single-block embedding in Eq.~\eqref{eq:app_sw_embedding_ico_sup},
with only the $\lambda$-th component equal to $x$, and define 
\begin{equation}
\mathcal{T}^{\lambda\to\nu}_{{\rm out}}(x):=\left[\operatorname{Tr}_{{\rm out}}\!\left(\iota_{\lambda}(x)\right)\right]^{{\rm in}}_{\nu},\;\lambda\in\mathbb{Y}^{(d^{2})}_{N},\;\nu\in\mathbb{Y}^{(d)}_{N}.\label{eq:partial_trace_reduced_map_symred}
\end{equation}
If $S=\iota_{{\rm SW}}(s)$, then Eq.~\eqref{eq:para_unreduced_causal_constraint_symred}
is equivalent blockwise to 
\begin{equation}
\mu d^{N}I_{\mathcal{U}^{{\rm in}}_{\nu}}-\sum_{\lambda\in\mathbb{Y}^{(d^{2})}_{N}}\mathcal{T}^{\lambda\to\nu}_{{\rm out}}(s_{\lambda})\succeq0,\;\nu\in\mathbb{Y}^{(d)}_{N}.\label{eq:para_reduced_causal_constraint_symred}
\end{equation}

Therefore, for $\mathsf{Bd}\in\{\mathsf{SLD},\mathsf{HN},\mathsf{NH}\}$,
the reduced dual problem for the Para strategy is 
\begin{equation}
\begin{aligned}\max\; & -\operatorname{Tr}[h]-\mu d^{N}\\
\mathrm{s.t.}\; & B^{\mathsf{Bd}}_{\lambda}\succeq0, &  & \forall\lambda\in\mathbb{Y}^{(d^{2})}_{N},\\
 & \mu d^{N}I_{\mathcal{U}^{{\rm in}}_{\nu}}-\sum_{\lambda\in\mathbb{Y}^{(d^{2})}_{N}}\mathcal{T}^{\lambda\to\nu}_{{\rm out}}(s_{\lambda})\succeq0, &  & \forall\nu\in\mathbb{Y}^{(d)}_{N}.
\end{aligned}
\label{eq:para_reduced_psd_symred}
\end{equation}
For the TCR bound, one only replaces the first group of constraints
by Eq.~\eqref{eq:tcr_measurement_reduced_symred}: 
\begin{equation}
\begin{aligned}\max\; & -\operatorname{Tr}[h]-\mu d^{N}\\
\mathrm{s.t.}\; & \left(B^{\mathsf{TCR}}_{\lambda}\right)_{\lambda\in\mathbb{Y}^{(d^{2})}_{N}}\in\mathcal{S}^{*,{\rm SR}}_{{\rm SEP}},\\
 & \mu d^{N}I_{\mathcal{U}^{{\rm in}}_{\nu}}-\sum_{\lambda\in\mathbb{Y}^{(d^{2})}_{N}}\mathcal{T}^{\lambda\to\nu}_{{\rm out}}(s_{\lambda})\succeq0, &  & \forall\nu\in\mathbb{Y}^{(d)}_{N}.
\end{aligned}
\label{eq:para_reduced_tcr_symred}
\end{equation}

\subsection{Sup strategy}

Let $e\in\mathfrak{S}_{N}$ be the identity permutation, and let $\mathcal{Q}^{(e)}_{\mathsf{FO}}$
be the dual linear map for the standard fixed-order comb. For any
$\pi\in\mathfrak{S}_{N}$, define the permuted fixed-order map by
\begin{equation}
\mathcal{Q}^{(\pi)}_{\mathsf{FO}}(X):=U_{\pi}\mathcal{Q}^{(e)}_{\mathsf{FO}}\!\left(U^{\dagger}_{\pi}XU_{\pi}\right)U^{\dagger}_{\pi}.\label{eq:app_permuted_fo_map}
\end{equation}

The unreduced $\mathsf{Sup}$ causal constraint requires the same
test operator $S$ to be upper-bounded by all fixed-order dual combs:
\begin{equation}
\mathcal{Q}^{(\pi)}_{\mathsf{FO}}(\Lambda_{\pi})+\mu I_{{\rm sys}}-S\succeq0,\;\forall\,\pi\in\mathfrak{S}_{N},\label{eq:app_sup_all_order_constraints}
\end{equation}
where $\Lambda_{\pi}=\Lambda^{\dagger}_{\pi}\in\mathcal{L}(\mathcal{H}_{{\rm sys}})$.
For any $\tau\in\mathfrak{S}_{N}$, let $S_{\tau}:=U_{\tau}SU^{\dagger}_{\tau},\;\Lambda^{(\tau)}_{\pi}:=U_{\tau}\Lambda_{\tau^{-1}\pi}U^{\dagger}_{\tau}.$
By Eq.~\eqref{eq:app_permuted_fo_map}, $\mathcal{Q}^{(\pi)}_{\mathsf{FO}}\!\left(\Lambda^{(\tau)}_{\pi}\right)=U_{\tau}\mathcal{Q}^{(\tau^{-1}\pi)}_{\mathsf{FO}}\!\left(\Lambda_{\tau^{-1}\pi}\right)U^{\dagger}_{\tau}.$
Thus, if $(S,\{\Lambda_{\pi}\}_{\pi\in\mathfrak{S}_{N}})$ satisfies
Eq.~\eqref{eq:app_sup_all_order_constraints}, then $(S_{\tau},\{\Lambda^{(\tau)}_{\pi}\}_{\pi\in\mathfrak{S}_{N}})$
is feasible as well. Combining the covariance of Eq.~\eqref{eq:app_sup_all_order_constraints}
with the group-averaging argument at the beginning of this section,
the strategy-dual variable can be restricted without loss to $S=\iota_{{\rm SW}}(s),$
and the corresponding precision-bound constraints are given by Eqs.~\eqref{eq:measurement_block_psd_symred}
and~\eqref{eq:tcr_measurement_reduced_symred}.

Once $S=\iota_{{\rm SW}}(s)$ is permutation invariant, the $N!$
order constraints in Eq.~\eqref{eq:app_sup_all_order_constraints}
reduce further to a single standard-order constraint. Suppose there
exists $\Lambda=\Lambda^{\dagger}\in\mathcal{L}(\mathcal{H}_{{\rm sys}})$
such that 
\begin{equation}
\mathcal{Q}^{(e)}_{\mathsf{FO}}(\Lambda)+\mu I_{{\rm sys}}-\iota_{{\rm SW}}(s)\succeq0,\label{eq:app_sup_standard_order_constraint}
\end{equation}
then, for any $\pi\in\mathfrak{S}_{N}$, set $\Lambda_{\pi}:=U_{\pi}\Lambda U^{\dagger}_{\pi}.$
Using Eq.~\eqref{eq:app_permuted_fo_map} and $U_{\pi}\iota_{{\rm SW}}(s)U^{\dagger}_{\pi}=\iota_{{\rm SW}}(s),$
we obtain $\mathcal{Q}^{(\pi)}_{\mathsf{FO}}(\Lambda_{\pi})+\mu I_{{\rm sys}}-\iota_{{\rm SW}}(s)\succeq0.$
Conversely, if Eq.~\eqref{eq:app_sup_all_order_constraints} holds
for all $\pi$, then choosing $\pi=e$ gives Eq.~\eqref{eq:app_sup_standard_order_constraint}.
Hence, once the strategy-dual variable is permutation invariant, all
fixed-order constraints are equivalent to the single standard-order
constraint.

Therefore, for $\mathsf{Bd}\in\{\mathsf{SLD},\mathsf{HN},\mathsf{NH}\}$,
the semireduced dual problem for the $\mathsf{Sup}$ strategy is 
\begin{equation}
\begin{aligned}\max\; & -\operatorname{Tr}[h]-\mu d^{N}\\
\mathrm{s.t.}\; & B^{\mathsf{Bd}}_{\lambda}\succeq0,\forall\,\lambda\in\mathbb{Y}^{(d^{2})}_{N},\\
 & \mathcal{Q}^{(e)}_{\mathsf{FO}}(\Lambda)+\mu I_{{\rm sys}}-\iota_{{\rm SW}}(s)\succeq0.
\end{aligned}
\label{eq:app_sup_reduced_psd}
\end{equation}
The TCR-bound case is 
\begin{equation}
\begin{aligned}\max\; & -\operatorname{Tr}[h]-\mu d^{N}\\
\mathrm{s.t.}\; & \left(B^{\mathsf{TCR}}_{\lambda}\right)_{\lambda\in\mathbb{Y}^{(d^{2})}_{N}}\in\mathcal{S}^{*,{\rm SR}}_{{\rm SEP}},\\
 & \mathcal{Q}^{(e)}_{\mathsf{FO}}(\Lambda)+\mu I_{{\rm sys}}-\iota_{{\rm SW}}(s)\succeq0.
\end{aligned}
\label{eq:app_sup_reduced_tcr}
\end{equation}

The standard-order map $\mathcal{Q}^{(e)}_{\mathsf{FO}}$ does not
in general satisfy $\mathcal{Q}^{(e)}_{\mathsf{FO}}\!\left(U_{\pi}XU^{\dagger}_{\pi}\right)=U_{\pi}\mathcal{Q}^{(e)}_{\mathsf{FO}}(X)U^{\dagger}_{\pi}.$
Therefore, one cannot further group-average the single standard-order
multiplier $\Lambda$. Imposing $[\Lambda,U_{\pi}]=0,\;\forall\,\pi\in\mathfrak{S}_{N},$
would artificially shrink the original dual feasible set.

\subsection{Gen strategy}

\label{app:ico_sup_symmetry_reduction}

The unreduced dual causal constraint for the Gen strategy is 
\begin{equation}
\mathcal{Q}_{\mathsf{Gen}}(\Lambda)+\mu I_{{\rm sys}}-S\succeq0,\label{eq:app_ico_full_causal_constraint}
\end{equation}
where $\Lambda=\Lambda^{\dagger}\in\mathcal{L}(\mathcal{H}_{{\rm sys}})$.

By Eq.~\eqref{eq:qico_appendix}, $\mathcal{Q}_{\mathsf{Gen}}$ is
covariant under permutations of the channel queries: $\mathcal{Q}_{\mathsf{Gen}}\!\left(U_{\pi}\Lambda U^{\dagger}_{\pi}\right)=U_{\pi}\mathcal{Q}_{\mathsf{Gen}}(\Lambda)U^{\dagger}_{\pi},\;\forall\,\pi\in\mathfrak{S}_{N}.$
Therefore, Eq.~\eqref{eq:app_ico_full_causal_constraint} remains
feasible when $S$ and $\Lambda$ are conjugated by the same permutation.
By the group-averaging argument at the beginning of this section,
we may assume without loss of generality that 
\begin{equation}
S=\iota_{{\rm SW}}(s),\;\Lambda=\iota_{{\rm SW}}(\ell),\label{eq:app_ico_invariant_variables}
\end{equation}
where $s=\{s_{\lambda}\}_{\lambda\in\mathbb{Y}^{(d^{2})}_{N}},\;\ell=\{\ell_{\nu}\}_{\nu\in\mathbb{Y}^{(d^{2})}_{N}},$
and $s_{\lambda}=s^{\dagger}_{\lambda}\in\mathcal{L}(\mathcal{U}_{\lambda}),\;\ell_{\nu}=\ell^{\dagger}_{\nu}\in\mathcal{L}(\mathcal{U}_{\nu}).$

To write the causal constraint blockwise, define 
\begin{equation}
\mathcal{Q}^{\nu\to\lambda}_{\mathsf{Gen}}(\ell_{\nu}):=\left[\mathcal{Q}_{\mathsf{Gen}}\!\left(\iota_{\nu}(\ell_{\nu})\right)\right]_{\lambda},\;\nu,\lambda\in\mathbb{Y}^{(d^{2})}_{N}.\label{eq:gen_reduced_projection_map}
\end{equation}
Since $\Lambda=\sum_{\nu\in\mathbb{Y}^{(d^{2})}_{N}}\iota_{\nu}(\ell_{\nu})$,
linearity gives that the $\lambda$-th Schur block of $\mathcal{Q}_{\mathsf{Gen}}(\Lambda)$
is $\left[\mathcal{Q}_{\mathsf{Gen}}(\Lambda)\right]_{\lambda}=\sum_{\nu\in\mathbb{Y}^{(d^{2})}_{N}}\mathcal{Q}^{\nu\to\lambda}_{\mathsf{Gen}}(\ell_{\nu}).$
Hence Eq.~\eqref{eq:app_ico_full_causal_constraint} is equivalent
to 
\begin{equation}
\sum_{\nu\in\mathbb{Y}^{(d^{2})}_{N}}\mathcal{Q}^{\nu\to\lambda}_{\mathsf{Gen}}(\ell_{\nu})+\mu I_{\mathcal{U}_{\lambda}}-s_{\lambda}\succeq0,\;\forall\,\lambda\in\mathbb{Y}^{(d^{2})}_{N}.\label{eq:app_ico_reduced_causal_constraint}
\end{equation}

Therefore, for $\mathsf{Bd}\in\{\mathsf{SLD},\mathsf{HN},\mathsf{NH}\}$,
the reduced dual problem for the $\mathsf{Gen}$ strategy is 
\begin{equation}
\begin{aligned}\max\; & -\operatorname{Tr}[h]-\mu d^{N}\\
\mathrm{s.t.}\; & B^{\mathsf{Bd}}_{\lambda}\succeq0, &  & \forall\,\lambda\in\mathbb{Y}^{(d^{2})}_{N},\\
 & \sum_{\nu\in\mathbb{Y}^{(d^{2})}_{N}}\mathcal{Q}^{\nu\to\lambda}_{\mathsf{Gen}}(\ell_{\nu})+\mu I_{\mathcal{U}_{\lambda}}-s_{\lambda}\succeq0, &  & \forall\,\lambda\in\mathbb{Y}^{(d^{2})}_{N}.
\end{aligned}
\label{eq:app_ico_reduced_psd}
\end{equation}
The TCR-bound case is 
\begin{equation}
\begin{aligned}\max\; & -\operatorname{Tr}[h]-\mu d^{N}\\
\mathrm{s.t.}\; & \left(B^{\mathsf{TCR}}_{\lambda}\right)_{\lambda\in\mathbb{Y}^{(d^{2})}_{N}}\in\mathcal{S}^{*,{\rm SR}}_{{\rm SEP}},\\
 & \sum_{\nu\in\mathbb{Y}^{(d^{2})}_{N}}\mathcal{Q}^{\nu\to\lambda}_{\mathsf{Gen}}(\ell_{\nu})+\mu I_{\mathcal{U}_{\lambda}}-s_{\lambda}\succeq0, &  & \forall\,\lambda\in\mathbb{Y}^{(d^{2})}_{N}.
\end{aligned}
\label{eq:app_ico_reduced_tcr}
\end{equation}

Conversely, any variables satisfying these block constraints can be
embedded back into the full space by $\iota_{{\rm SW}}$, yielding
permutation-invariant variables $S$ and $\Lambda$ that satisfy Eq.~\eqref{eq:app_ico_full_causal_constraint}.
The precision-bound constraints are also recovered from Eq.~\eqref{eq:measurement_block_psd_symred}
or Eq.~\eqref{eq:tcr_measurement_reduced_symred}, and the embedding
does not change the objective value. Hence the reduced and unreduced
problems have the same optimum: 
\begin{equation}
S(D^{\mathsf{Gen}}_{\mathsf{Bd}})=S(D^{\mathsf{Gen,SR}}_{\mathsf{Bd}}),\;\mathsf{Bd}\in\{\mathsf{SLD},\mathsf{HN},\mathsf{NH},\mathsf{TCR}\}.\label{eq:app_ico_value_equivalence}
\end{equation}

\section{Doherty--Parrilo--Spedalieri (DPS) hierarchy and symmetry reduction}

\label{app:dps-symmetry-reduction}

This section gives the dual problem for the DPS relaxation and the
symmetry-reduced form used in the numerical calculations. The dual
problem $S(D^{\mathsf{Strat}}_{\mathsf{DPS}_{k}})$ can be written
as 
\begin{equation}
\begin{aligned}\underset{h,\mu,S,\{Y_{uv}\},\{Z_{m}\}}{\mathrm{maximize}}\; & -\operatorname{Tr}[h]-\mu d^N\\
\mathrm{s.t.}\; & \Gamma_{{\rm full}}\succeq0,\;Z_{m}\succeq0,\;m=1,\ldots,k,\\
 & \mu I_{{\rm sys}}-S\in\mathcal{W}^{*}_{\mathsf{Strat}}.
\end{aligned}
\end{equation}
Here the dual slack operator is 
\begin{equation}
\begin{aligned}
\Gamma_{{\rm full}}={}&\;G\otimes I^{\otimes(k-1)}_{\mathcal{R}}\otimes W_{\theta}\\
&+\sum^{t}_{i,j=1}h_{ij}\left(K_{i}\otimes I^{\otimes(k-1)}_{\mathcal{R}}\otimes\partial_{j}W_{\theta}\right)\\
&+\sum_{0\leq u<v\leq t}\left(R_{uv}\otimes I^{\otimes(k-1)}_{\mathcal{R}}\otimes Y_{uv}\right)\\
&+K_{0}\otimes I^{\otimes(k-1)}_{\mathcal{R}}\otimes S-\sum^{k}_{m=1}Z^{T_{\mathcal{R}^{\otimes m}}}_{m}.
\end{aligned}
\end{equation}
Here $S=S^{\dagger}$ is the free Hermitian variable used in the unified
dual, and $\mu I_{{\rm sys}}-S\in\mathcal{W}^{*}_{\mathsf{Strat}}$
means that $\mu I_{{\rm sys}}-S$ belongs to the dual cone of the
chosen strategy cone.

Without loss of generality, one may restrict the problem directly
to the fully symmetric subspace $\mathrm{Sym}^{k}(\mathcal{R})$~\cite{doherty2004CompleteFamilySeparability},
where $\mathrm{Sym}^{k}(\mathcal{R})\subset\mathcal{R}^{\otimes k}$
is the linear span of symmetrized states. Let ${\displaystyle E_{k}:\mathrm{Sym}^{k}(\mathcal{R})\hookrightarrow\mathcal{R}^{\otimes k}}$
denote the embedding of the fully symmetric subspace into the full
tensor-product space. For any single-copy operator $B\in\mathcal{L}(\mathcal{R})$,
define the symmetrized operator ${\displaystyle \mathcal{M}_{k}(B):=\frac{1}{k}E^{\dagger}_{k}\left(\sum^{k}_{r=1}B^{(r)}\right)E_{k},}$
where $B^{(r)}$ means that $B$ acts on the $r$-th copy of $\mathcal{R}$
and the identity acts on all other copies.

To write the reduced positive-partial-transpose (PPT) conditions,
let ${\displaystyle E_{m,k-m}:=E_{m}\otimes E_{k-m}.}$ For ${\displaystyle \widetilde{x}_{k}\in\mathcal{L}\left(\mathrm{Sym}^{k}(\mathcal{R})\otimes\mathcal{H}_{{\rm sys}}\right),}$
define the reduced PPT map 
\begin{equation}
\begin{aligned}
\mathcal{P}_{m}(\widetilde{x}_{k}):={}&\left(E^{\dagger}_{m,k-m}\otimes I_{{\rm sys}}\right)\\
&\times\Bigl[\left(E_{k}\otimes I_{{\rm sys}}\right)\widetilde{x}_{k}
\left(E^{\dagger}_{k}\otimes I_{{\rm sys}}\right)\Bigr]^{T_{\mathcal{R}^{\otimes m}}}\\
&\times\left(E_{m,k-m}\otimes I_{{\rm sys}}\right).
\end{aligned}
\end{equation}
The output space of $\mathcal{P}_{m}$ is
\begin{equation*}
\begin{aligned}
&\mathcal{L}\bigl(\mathrm{Sym}^{m}(\mathcal{R})\\
&\qquad\otimes\mathrm{Sym}^{k-m}(\mathcal{R})\otimes\mathcal{H}_{{\rm sys}}\bigr).
\end{aligned}
\end{equation*}
The corresponding adjoint map is 
\begin{equation}
\begin{aligned}
\mathcal{P}^{\dagger}_{m}(Z_{m}):={}&\left(E^{\dagger}_{k}\otimes I_{{\rm sys}}\right)\\
&\times\Bigl[\left(E_{m,k-m}\otimes I_{{\rm sys}}\right)Z_{m}\\
&\qquad\times\left(E^{\dagger}_{m,k-m}\otimes I_{{\rm sys}}\right)\Bigr]^{T_{\mathcal{R}^{\otimes m}}}\\
&\times\left(E_{k}\otimes I_{{\rm sys}}\right).
\end{aligned}
\end{equation}

With this notation, the $\mathcal{R}$-reduced primal semidefinite
program (SDP) $S(P^{\mathsf{Strat}}_{\mathrm{Sym}\text{-}\mathrm{DPS}_{k}})$
is 
\begin{equation}
\begin{aligned}\underset{\widetilde{x}_{k}}{\mathrm{minimize}}\; & \operatorname{Tr}\left[\left(\mathcal{M}_{k}(G)\otimes W_{\theta}\right)\widetilde{x}_{k}\right]\\
\mathrm{s.t.}\; & \operatorname{Tr}\left[\left(\mathcal{M}_{k}(K_{i})\otimes\partial_{j}W_{\theta}\right)\widetilde{x}_{k}\right]\\
&=\delta_{ij},\quad i,j=1,\ldots,t,\\
 & \operatorname{Tr}_{\mathrm{Sym}^{k}(\mathcal{R})}\left[\left(\mathcal{M}_{k}(K_{0})\otimes I_{{\rm sys}}\right)\widetilde{x}_{k}\right]\\
&\qquad\in\mathcal{W}_{\mathsf{Strat}},\\
 & \operatorname{Tr}\left[\left(\mathcal{M}_{k}(K_{0})\otimes I_{{\rm sys}}\right)\widetilde{x}_{k}\right]=d^N,\\
 & \operatorname{Tr}_{\mathrm{Sym}^{k}(\mathcal{R})}\left[\left(\mathcal{M}_{k}(R_{uv})\otimes I_{{\rm sys}}\right)\widetilde{x}_{k}\right]\\
&\qquad=0,\quad 0\leq u<v\leq t,\\
 & \widetilde{x}_{k}\succeq0,\;\mathcal{P}_{m}(\widetilde{x}_{k})\succeq0,\;m=1,\ldots,k.
\end{aligned}
\label{eq:app-R-reduced-primal}
\end{equation}
Here $\widetilde{x}_{k}$ already lives on $\mathrm{Sym}^{k}(\mathcal{R})\otimes\mathcal{H}_{{\rm sys}}$
so no explicit full-symmetry projection constraint is needed.

The corresponding dual problem $S(D^{\mathsf{Strat}}_{\mathrm{Sym}\text{-}DPS_{k}})$
is 
\begin{equation}
\begin{aligned}\underset{h,\mu,S,\{Y_{uv}\},\{Z_{m}\}}{\mathrm{maximize}}\; & -\operatorname{Tr}[h]-\mu d^N\\
\mathrm{s.t.}\; & \Gamma_{\mathcal{R}}\succeq0,Z_{m}\succeq0,\;m=1,\ldots,k,\\
 & \mu I_{{\rm sys}}-S\in\mathcal{W}^{*}_{\mathsf{Strat}}.
\end{aligned}
\label{eq:app-R-reduced-dual}
\end{equation}
Here 
\begin{equation}
\begin{aligned}\Gamma_{\mathcal{R}}={}&\mathcal{M}_{k}(G)\otimes W_{\theta}\\
&+\sum^{t}_{i,j=1}h_{ij}\left[\mathcal{M}_{k}(K_{i})\otimes\partial_{j}W_{\theta}\right]\\
&+\sum_{0\leq u<v\leq t}\left[\mathcal{M}_{k}(R_{uv})\otimes Y_{uv}\right]\\
&+\mathcal{M}_{k}(K_{0})\otimes S-\sum^{k}_{m=1}\mathcal{P}^{\dagger}_{m}(Z_{m}).
\end{aligned}
\end{equation}
The variables $Z_{m}$ satisfy
\begin{equation*}
\begin{aligned}
Z_{m}\in{}&\mathcal{L}\bigl(\mathrm{Sym}^{m}(\mathcal{R})\\
&\qquad\otimes\mathrm{Sym}^{k-m}(\mathcal{R})\otimes\mathcal{H}_{{\rm sys}}\bigr).
\end{aligned}
\end{equation*}

Equations~\eqref{eq:app-R-reduced-primal} and~\eqref{eq:app-R-reduced-dual}
are the basic forms used below. These two forms compress only the
DPS auxiliary system $\mathcal{R}$ and do not yet use the permutation
symmetry of the physical system $\mathcal{H}_{{\rm sys}}$.

\subsection{Double symmetry reduction for the Para strategy}

\label{app:para-double-reduction}

Equation~\eqref{eq:app-R-reduced-dual} has already used the permutation
symmetry of the $k$-th DPS extension on the auxiliary system $\mathcal{R}^{\otimes k}$.
We now further use the permutation symmetry of the $N$ channel queries
in the Para strategy on the physical system $\mathcal{H}_{{\rm sys}}=\left(\mathcal{H}_{{\rm in}}\otimes\mathcal{H}_{{\rm out}}\right)^{\otimes N}$
to reduce the problem size. By the preceding group-averaging argument,
the physical-system dual variables in Eq.~\eqref{eq:app-R-reduced-dual}
can be chosen permutation invariant. Hence there exist Hermitian blocks
$s_{\lambda},Y_{uv,\lambda}\in\mathcal{L}(\mathcal{U}_{\lambda}),\;\lambda\in\mathbb{Y}^{(d^{2})}_{N},\;0\leq u<v\leq t,$
such that 
\begin{equation}
\begin{aligned}\mathsf{U}S\mathsf{U}^{\dagger} & =\bigoplus_{\lambda\in\mathbb{Y}^{(d^{2})}_{N}}s_{\lambda}\otimes I_{\mathcal{S}_{\lambda}},\\
\mathsf{U}Y_{uv}\mathsf{U}^{\dagger} & =\bigoplus_{\lambda\in\mathbb{Y}^{(d^{2})}_{N}}Y_{uv,\lambda}\otimes I_{\mathcal{S}_{\lambda}}.
\end{aligned}
\label{eq:app-S-schur-blocks}
\end{equation}
Here $t$ is the number of parameters, and $S$ and $Y_{uv}$ are
respectively the free Hermitian variable in the unified dual and the
real-estimator constraint multipliers in Eq.~\eqref{eq:app-R-reduced-dual}.

For each $m=1,\ldots,k$, the PPT dual variable defined above, $Z_{m}\in\mathcal{L}\!\left(\operatorname{Sym}^{m}(\mathcal{R})\otimes\operatorname{Sym}^{k-m}(\mathcal{R})\otimes\mathcal{H}_{{\rm sys}}\right)$
can likewise be chosen permutation invariant on the physical system:
\begin{equation}
\begin{aligned}
&\left(I_{\mathrm{Sym}^{m}(\mathcal{R})\otimes\mathrm{Sym}^{k-m}(\mathcal{R})}\otimes\mathsf{U}\right)\\
&\quad\times Z_{m}\left(I_{\mathrm{Sym}^{m}(\mathcal{R})\otimes\mathrm{Sym}^{k-m}(\mathcal{R})}\otimes\mathsf{U}^{\dagger}\right)\\
&=\bigoplus_{\lambda\in\mathbb{Y}^{(d^{2})}_{N}}Z_{m,\lambda}\otimes I_{\mathcal{S}_{\lambda}},
\end{aligned}
\label{eq:app-Zm-schur-blocks}
\end{equation}
where $Z_{m,\lambda}\succeq0,\;Z_{m,\lambda}\in\mathcal{L}\!\left(\operatorname{Sym}^{m}(\mathcal{R})\otimes\operatorname{Sym}^{k-m}(\mathcal{R})\otimes\mathcal{U}_{\lambda}\right).$

Use the output partial-trace block map defined in Eq.~\eqref{eq:partial_trace_reduced_map_symred},
$\mathcal{T}^{\lambda\to\nu}_{{\rm out}}:\mathcal{L}(\mathcal{U}_{\lambda})\longrightarrow\mathcal{L}(\mathcal{U}^{{\rm in}}_{\nu}),\;\lambda\in\mathbb{Y}^{(d^{2})}_{N},\;\nu\in\mathbb{Y}^{(d)}_{N}.$
By the characterization of the Para dual cone $\mathcal{W}^{*}_{\mathsf{Para}}$,
$\mu I_{{\rm sys}}-S\in\mathcal{W}^{*}_{\mathsf{Para}}$ is equivalent
to $\mu d^{N}I_{{\rm in}}-\operatorname{Tr}_{{\rm out}}S\succeq0$,
and hence to 
\begin{equation}
\mu d^{N}I_{\mathcal{U}^{{\rm in}}_{\nu}}-\sum_{\lambda\in\mathbb{Y}^{(d^{2})}_{N}}\mathcal{T}^{\lambda\to\nu}_{{\rm out}}(s_{\lambda})\succeq0,\;\forall\,\nu\in\mathbb{Y}^{(d)}_{N}.\label{eq:app-para-dps-causal-block}
\end{equation}

Let $\mathcal{P}^{\dagger}_{m}$ also denote its natural extension
to each physical Schur block: 
\begin{equation}
\mathcal{P}^{\dagger}_{m}\big|_{\lambda}:=\mathcal{P}^{\dagger}_{m,\mathcal{R}}\otimes\operatorname{id}_{\mathcal{L}(\mathcal{U}_{\lambda})},\label{eq:app-Pm-block-extension}
\end{equation}
where $\mathcal{P}^{\dagger}_{m,\mathcal{R}}$ denotes the action
of $\mathcal{P}^{\dagger}_{m}$ on the DPS auxiliary factor, and $\operatorname{id}_{\mathcal{L}(\mathcal{U}_{\lambda})}$
is the identity superoperator on $\mathcal{L}(\mathcal{U}_{\lambda})$.
To simplify notation, below we still write $\left(\mathcal{P}^{\dagger}_{m,\mathcal{R}}\otimes\operatorname{id}_{\mathcal{L}(\mathcal{U}_{\lambda})}\right)[Z_{m,\lambda}]$
as $\mathcal{P}^{\dagger}_{m}(Z_{m,\lambda})$.

Substituting these block forms into Eq.~\eqref{eq:app-R-reduced-dual}
gives the double symmetry-reduced form of the $k$-th DPS dual problem
for the Para strategy: 
\begin{equation}
\begin{aligned}\underset{\substack{h,\mu,\{s_{\lambda}\},\{Y_{uv,\lambda}\},\{Z_{m,\lambda}\}}
}{\mathrm{maximize}}\! & -\operatorname{Tr}[h]-\mu d^{N}\\
\mathrm{s.t.} & \Gamma^{\mathsf{Para}}_{\lambda}\succeq0,\forall\,\lambda\in\mathbb{Y}^{(d^{2})}_{N},\\
 & Z_{m,\lambda}\succeq0,m=1,\ldots,k,\;\forall\,\lambda\in\mathbb{Y}^{(d^{2})}_{N},\\
 & \mu d^{N}I_{\mathcal{U}^{{\rm in}}_{\nu}}-\sum_{\lambda\in\mathbb{Y}^{(d^{2})}_{N}}\mathcal{T}^{\lambda\to\nu}_{{\rm out}}(s_{\lambda})\\
&\qquad\succeq0,\;\forall\,\nu\in\mathbb{Y}^{(d)}_{N}.
\end{aligned}
\label{eq:app-para-double-reduced-dual}
\end{equation}
Here $h\in\mathbb{R}^{t\times t},\;\mu\in\mathbb{R},$ and the LMI
operator on each physical Schur block is 
\begin{equation}
\begin{aligned}
\Gamma^{\mathsf{Para}}_{\lambda}:={}&\mathcal{M}_{k}(G)\otimes W_{\theta,\lambda}\\
&+\sum^{t}_{i,j=1}h_{ij}\,\mathcal{M}_{k}(K_{i})\otimes\partial_{j}W_{\theta,\lambda}\\
&+\sum_{0\leq u<v\leq t}\mathcal{M}_{k}(R_{uv})\otimes Y_{uv,\lambda}\\
&+\mathcal{M}_{k}(K_{0})\otimes s_{\lambda}-\sum^{k}_{m=1}\mathcal{P}^{\dagger}_{m}(Z_{m,\lambda}),\\
&\qquad\lambda\in\mathbb{Y}^{(d^{2})}_{N}.
\end{aligned}
\label{eq:app-para-block-LMI}
\end{equation}

Equation~\eqref{eq:app-para-double-reduced-dual} uses both the permutation
symmetry of the DPS auxiliary copies in $\mathcal{R}^{\otimes k}$
and the permutation symmetry of the $N$ channel queries in $\mathcal{H}_{{\rm sys}}$.
The block form decomposes the original dual problem into smaller semidefinite
constraints labeled by $\lambda\in\mathbb{Y}^{(d^{2})}_{N}$. Other
strategies can be reduced similarly by using their own strategy dual
cones, so we do not spell them out here.

\section{Symmetry reduction of the AP method}

\label{app:ap-sym-reduction}

This section gives the system-side reduction, based on Schur--Weyl
duality, of the AP inner approximation in Eq.~\eqref{eq:tcr_sandwich}
of the main text for the Para strategy. We use the Young diagrams,
Schur transform, and block-extraction notation of Eq.~\eqref{eq:schur_weyl}
in the main text and Sec.~\ref{app:schur-weyl}. Specifically, $[X]_{\lambda}$
and $\iota_{{\rm SW}}$ are defined in Eqs.~\eqref{eq:schur_block_extraction_symred}
and~\eqref{eq:app_sw_embedding_ico_sup}, and $d_{\lambda}=\dim\mathcal{S}_{\lambda}$.

We use the operators $K_{i}$ defined after Eq.~\eqref{eq:equivalence_forward_pairing},
with $K_{0}=|0\rangle\langle0|$. For auxiliary-side vertices $\mathcal{V}_{A}=\{\sigma_{\alpha}\}^{M_{A}}_{\alpha=1}$,
set $p_{\alpha}=\operatorname{Tr}(K_{0}\sigma_{\alpha})$, $c^{i}_{\alpha}=\operatorname{Tr}(K_{i}\sigma_{\alpha})$,
and $g_{\alpha}=\operatorname{Tr}(G\sigma_{\alpha})$. The system-side
variables are written as $\widetilde{\tau}_{\alpha}=\iota_{{\rm SW}}((\tau_{\alpha,\lambda})_{\lambda})$,
where $\tau_{\alpha,\lambda}\succeq0$.

\subsection{Block constraints for the Para strategy}

\label{app:ap-para-reduction}

By Eq.~\eqref{eq:cone_para}, a Para process satisfies $P=\rho_{{\rm in}}\otimes I^{\otimes N}_{{\rm out}}$.
Write the permutation-invariant $\rho_{{\rm in}}$ in input Schur
blocks $\{\rho_{\nu}\}_{\nu\in\mathbb{Y}^{(d)}_{N}}$, and let $\mathcal{J}^{\nu\to\lambda}_{{\rm out}}$
be the linear map induced by $\rho_{{\rm in}}\mapsto\rho_{{\rm in}}\otimes I^{\otimes N}_{{\rm out}}$
between the input block $\nu$ and the global block $\lambda$. The
Para constraint is then written blockwise as $x_{\lambda}=\sum_{\nu\in\mathbb{Y}^{(d)}_{N}}\mathcal{J}^{\nu\to\lambda}_{{\rm out}}(\rho_{\nu})$,
where $x_{\lambda}=\sum^{M_{A}}_{\alpha=1}p_{\alpha}\tau_{\alpha,\lambda}$
and $\rho_{\nu}\succeq0$.

The output partial-trace block map $\mathcal{T}^{\lambda\to\nu}_{{\rm out}}$
is defined in Eq.~\eqref{eq:partial_trace_reduced_map_symred}. The
block condition for the Para dual cone is given in Eq.~\eqref{eq:para_reduced_causal_constraint_symred}.

\subsection{Reduced $B$-step}

\label{app:ap-para-bob-reduced}

With $\mathcal{V}_{A}$ fixed, the reduced $B$-step is 
\begin{equation}
\begin{aligned}\min_{\{\tau_{\alpha,\lambda}\},\{\rho_{\nu}\}}\  & \sum^{M_{A}}_{\alpha=1}\sum_{\lambda\in\mathbb{Y}^{(d^{2})}_{N}}d_{\lambda}g_{\alpha}\operatorname{Tr}[W_{\theta,\lambda}\tau_{\alpha,\lambda}]\\
\mathrm{s.t.}\  & \sum^{M_{A}}_{\alpha=1}\sum_{\lambda\in\mathbb{Y}^{(d^{2})}_{N}}d_{\lambda}c^{i}_{\alpha}\operatorname{Tr}[\partial_{j}W_{\theta,\lambda}\tau_{\alpha,\lambda}]\\
&\qquad=\delta_{ij}\ \text{for}\ i,j=1,\ldots,t,\\
 & \sum^{M_{A}}_{\alpha=1}p_{\alpha}\tau_{\alpha,\lambda}\\
&\qquad=\sum_{\nu\in\mathbb{Y}^{(d)}_{N}}\mathcal{J}^{\nu\to\lambda}_{{\rm out}}(\rho_{\nu})\quad\text{for all }\lambda\in\mathbb{Y}^{(d^{2})}_{N},\\
 & \sum_{\lambda\in\mathbb{Y}^{(d^{2})}_{N}}d_{\lambda}\operatorname{Tr}\left[\sum^{M_{A}}_{\alpha=1}p_{\alpha}\tau_{\alpha,\lambda}\right]=d^N,\\
 & \tau_{\alpha,\lambda}\succeq0,\rho_{\nu}\succeq0.
\end{aligned}
\label{eq:app-ap-B-primal-para-SR}
\end{equation}

Let $h\in\mathbb{R}^{t\times t}$, $\mu\in\mathbb{R}$, and $s_{\lambda}=s^{\dagger}_{\lambda}$.
The dual is 
\begin{equation}
\begin{aligned}\max_{h,\mu,\{s_{\lambda}\}}\  & -\operatorname{Tr}h-\mu d^N\\
\mathrm{s.t.}\  & g_{\alpha}W_{\theta,\lambda}+\sum^{t}_{i,j=1}h_{ij}c^{i}_{\alpha}\partial_{j}W_{\theta,\lambda}+p_{\alpha}(\mu I_{\lambda}-s_{\lambda})\succeq0\\
 & \text{for all}\ \alpha=1,\ldots,M_{A},\ \lambda\in\mathbb{Y}^{(d^{2})}_{N},\\
 & \sum_{\lambda\in\mathbb{Y}^{(d^{2})}_{N}}\mathcal{T}^{\lambda\to\nu}_{{\rm out}}(s_{\lambda})\succeq0\ \text{for all}\ \nu\in\mathbb{Y}^{(d)}_{N}.
\end{aligned}
\label{eq:app-ap-B-dual-para-SR}
\end{equation}
Here $I_{\lambda}$ is the identity operator on $\mathcal{U}_{\lambda}$.

\subsection{Reduced $A$-step}

\label{app:ap-para-alice-reduced}

Fix system-side vertices $\tau_{\beta}=\iota_{{\rm SW}}((\tau_{\beta,\lambda})_{\lambda})$,
where $\tau_{\beta,\lambda}\succeq0$. Define $w_{\beta}=\sum_{\lambda\in\mathbb{Y}^{(d^{2})}_{N}}d_{\lambda}\operatorname{Tr}[W_{\theta,\lambda}\tau_{\beta,\lambda}]$
and $d^{j}_{\beta}=\sum_{\lambda\in\mathbb{Y}^{(d^{2})}_{N}}d_{\lambda}\operatorname{Tr}[\partial_{j}W_{\theta,\lambda}\tau_{\beta,\lambda}]$.
The reduced $A$-step is 
\begin{equation}
\begin{aligned}\min_{\{\widetilde{\sigma}_{\beta}\},\{\rho_{\nu}\}}\  & \sum^{M_{B}}_{\beta=1}w_{\beta}\operatorname{Tr}[G\widetilde{\sigma}_{\beta}]\\
\mathrm{s.t.}\  & \sum^{M_{B}}_{\beta=1}d^{j}_{\beta}\operatorname{Tr}[K_{i}\widetilde{\sigma}_{\beta}]\\
&\qquad=\delta_{ij}\ \text{for}\ i,j=1,\ldots,t,\\
 & \sum^{M_{B}}_{\beta=1}\operatorname{Tr}[K_{0}\widetilde{\sigma}_{\beta}]\tau_{\beta,\lambda}\\
&\qquad=\sum_{\nu\in\mathbb{Y}^{(d)}_{N}}\mathcal{J}^{\nu\to\lambda}_{{\rm out}}(\rho_{\nu})\quad\text{for all }\lambda\in\mathbb{Y}^{(d^{2})}_{N},\\
 & \sum_{\lambda\in\mathbb{Y}^{(d^{2})}_{N}}d_{\lambda}\operatorname{Tr}\left[\sum^{M_{B}}_{\beta=1}\operatorname{Tr}[K_{0}\widetilde{\sigma}_{\beta}]\tau_{\beta,\lambda}\right]\\
&\qquad=d^N,\\
 & \widetilde{\sigma}_{\beta}\succeq0,\widetilde{\sigma}_{\beta}=\widetilde{\sigma}^{\mathsf{T}}_{\beta},\rho_{\nu}\succeq0.
\end{aligned}
\label{eq:app-ap-A-primal-para-SR}
\end{equation}

The corresponding dual is 
\begin{equation}
\begin{aligned}\max_{h,\mu,\{s_{\lambda}\}}\  & -\operatorname{Tr}h-\mu d^N\\
\mathrm{s.t.}\  & w_{\beta}G+\sum^{t}_{i,j=1}h_{ij}d^{j}_{\beta}K_{i}\\
&\qquad+\left[\sum_{\lambda\in\mathbb{Y}^{(d^{2})}_{N}}d_{\lambda}\operatorname{Tr}[(\mu I_{\lambda}-s_{\lambda})\tau_{\beta,\lambda}]\right]K_{0}\succeq0\\
 & \text{for all}\ \beta=1,\ldots,M_{B},\\
 & \sum_{\lambda\in\mathbb{Y}^{(d^{2})}_{N}}\mathcal{T}^{\lambda\to\nu}_{{\rm out}}(s_{\lambda})\succeq0\ \text{for all}\ \nu\in\mathbb{Y}^{(d)}_{N}.
\end{aligned}
\label{eq:app-ap-A-dual-para-SR}
\end{equation}

Equations~\eqref{eq:app-ap-B-primal-para-SR}--\eqref{eq:app-ap-A-dual-para-SR}
give the system-side block form of the AP method for the Para strategy.
Other strategies can be treated in the same way only when the corresponding
strategy cone is compatible with query permutations. We do not repeat
the derivation.

\bibliography{references}

\end{document}